\documentclass[twocolumn,trackchanges]{aastex63}

\newenvironment{changemargin}[2]{%
\begin{list}{}{%
\setlength{\topsep}{0pt}%
\setlength{\textheight}{#1}%
\setlength{\topmargin}{#2}%
\setlength{\listparindent}{\parindent}%
\setlength{\itemindent}{\parindent}%
\setlength{\parsep}{\parskip}%
}%
\item[]}{\end{list}}

\usepackage{amsmath}

\newcommand{\dd}{{\rm d}}

\received{March 12, 2021}
\revised{May 24, 2021}
\accepted{June 6, 2021}
\published{September 7, 2021}

\submitjournal{ApJS}

\shorttitle{Stellar mass and SFR within a billion light-years}
\shortauthors{Biteau}

\begin{document}

\title{Stellar Mass and Star Formation Rate within a Billion Light-Years}

\correspondingauthor{Jonathan Biteau}
\email{biteau@in2p3.fr}

\author[0000-0002-4202-8939]{Jonathan Biteau}
\affiliation{Universit\'e Paris-Saclay, CNRS/IN2P3, IJCLab, 91405 Orsay, France}

\begin{abstract}
To develop galaxy-targeting approaches, the gravitational-wave community built a catalog of stellar mass in the local universe based on the 2MASS spectroscopic and photometric redshift surveys. By cleaning and supplementing this catalog, the present work aims to establish a near-infrared flux-limited sample to map both stellar mass and star formation rate (SFR) over the full sky. The 2MASS spectroscopic and photometric redshift surveys are cross-matched with the HyperLEDA database and the Local Volume sample at $d<11\,$Mpc, providing a flux-limited sample with revised distance estimates and corrections for incompleteness out to 350\,Mpc. Scaling relations with stellar mass as a function of morphology are used to construct a SFR cosmography in the local universe. Stellar mass and SFR densities converge towards values compatible with deep-field observations beyond 100\,Mpc. The 3D distribution of these two tracers is consistent with the distribution of matter deduced from cosmic flows. With spectroscopic redshifts available for about half of the ${\sim}\,$400,000 galaxies within 350\,Mpc and photometric distances with a 12\% uncertainty available for the other half, the present sample may find applications in both cosmology and astroparticle physics. The present work provides in particular new bases for modeling the large- and intermediate-scale anisotropies observed at ultra-high energies. The distribution of magnetic fields at Mpc scales, which can be deduced from the 3D distribution of matter, is inferred to be crucial in shaping the ultra-high-energy sky.
\end{abstract}

\keywords{galaxy masses --- large-scale structure of the universe --- particle astrophysics --- scaling relations --- sky surveys}

\section{Introduction} \label{sec:intro}

The brightness of the extragalactic sky is determined by the 3D distribution of luminous matter, from the outskirts of the Milky Way to the most distant voids and superclusters. Such 3D mappings have been achieved through photometric and spectroscopic observations, accessing not only the fluxes of the galaxies but also the projection of their cosmological and peculiar velocities along the line of sight. Distances of nearby galaxies ($z\lesssim0.01$), for which peculiar motion dominates over recession in the Hubble flow, are constrained through cosmic-ladder measurements. Observations of stars at the tip of the red giant branch by the Hubble Space Telescope \citep[HST, see][]{2019ApJ...882...34F} have in particular enabled tremendous progress in mapping the local universe over the past decade \citep[e.g.][]{2018AN....339..615K}.

The reference all-sky catalog of galaxies within the Hubble flow is the 2MASS Redshift Survey \citep[2MRS,][]{2012ApJS..199...26H}. Through near-infrared (NIR) photometry, a known proxy for stellar mass \citep[e.g.][]{2003ApJS..149..289B}, and spectroscopic distance measurements of about 45,000 galaxies, the 2MRS offered new avenues for cosmology. Applications range from the cataloging of groups, clusters, superclusters and voids out to ${\sim}\,140\,$Mpc \citep[e.g.][]{2010ApJ...709..483L, 2015AJ....149..171T}, constraints on structure growth and cosmological parameters \citep[e.g.][]{2010ApJ...709..483L, 2015MNRAS.450..317C, 2012MNRAS.424..472B, 2017MNRAS.470..445N, 2021arXiv210207291L}, to tests of the Copernican principle \citep[e.g.][]{2012MNRAS.427.1994G}, that is in particular the isotropy and spatial homogeneity of the universe on large scales. Astroparticle physics also benefited from such 3D mappings of stellar mass. The 2MRS was used to constrain the dark-matter decay component of the extragalactic gamma-ray background \citep[e.g.][]{2014PhRvD..90b3514A} and to bound the density of sources of astrophysical neutrinos at TeV-PeV energies \citep[e.g.][]{2020JCAP...07..042A} as well as that of ultra-high energy cosmic rays (UHECRs) beyond EeV energies \citep[e.g.][]{2013JCAP...05..009P}.\footnote{$1{\rm \,TeV} \equiv 10^{12}{\rm \,eV}$, $1{\rm \,PeV} \equiv 10^{15}{\rm \,eV}$, $1{\rm \,EeV} \equiv 10^{18}{\rm \,eV}$.}

UHECRs are the most energetic particles known in the universe and their origin remains elusive \citep[see e.g.][for a review]{2019FrASS...6...23B}. They have long been thought to be of extragalactic origin above the ankle in the cosmic-ray spectrum, located at $5.0 \pm 0.1 \pm 0.8\,$EeV \citep{2020PhRvD.102f2005A}. The ankle indeed marks a hardening of the UHECR flux as a function of energy, that is a likely distinct origin from the lower-energy spectral component. Moreover, Galactic sources are not expected to be able to confine UHECRs long enough for them to reach multi-EeV energies \citep{2020PhR...872....1B}. The discovery of a dipolar component beyond 8\,EeV by the Pierre Auger Collaboration has provided the first observational confirmation of the extragalactic origin of UHECRs beyond the ankle, with a UHECR dipole direction consistent with that inferred from 2MRS \cite[flux weighted dipole from][]{2006MNRAS.373...45E} after accounting for deflections in the Galactic magnetic field \citep{2017Sci...357.1266P, 2018ApJ...868....4A}. UHECRs are indeed charged atomic nuclei and, despite their tremendous energies and thus magnetic rigidities, they can be deflected or even confined by magnetic fields encountered across their journey. As demonstrated by measurements of the maximum atmospheric depths of extensive air showers \citep{Aab:2017cgk,2019arXiv190909073T}, the UHECR flux shows an increasingly heavy composition (nuclear mass higher than that of ionized hydrogen) with increasing energy beyond the ankle. The UHECR spectral and composition data fit within a scenario where intermediate mass nuclei are accelerated up to $5\times Z\,$EeV \citep{2017JCAP...04..038A, 2020PhRvL.125l1106A}, where $Z$ is the UHECR charge. In this best-fit scenario, the region between the ankle and the new feature detected by \cite{2020PhRvL.125l1106A} at $13 \pm 1 \pm 2 \,$EeV, hereafter the ``instep'', would then be dominated by the by-products of He nuclei, while the region between the instep and the toe in the UHECR spectrum, located at $46 \pm 3 \pm 6 \,$EeV, would be dominated by the by-products of CNO nuclei. 

One of the difficulties in identifying the sources of UHECRs lies in the strong variation of their lifetime with energy. He and CNO nuclei in particular undergo photo-dissociation on photons from the cosmic microwave and infrared backgrounds after a travel time of about one Gyr at energies of ${\sim}\,15\,$EeV and ${\sim}\,10\,$EeV, or of about 10\,Myrs at energies of ${\sim}\,45\,$EeV and ${\sim}\,70\,$EeV, respectively \citep{2013APh....41...94A}. Such variations of lifetime by two orders of magnitude over only half a decade in energy could explain why the distribution of UHECR arrival directions appears to be consistent with the 2MRS dipole above the ankle, while possibly being correlated with the distribution of galaxies within a few Mpc around the toe \citep{2018ApJ...853L..29A}. This tentative correlation is supported by a ${\sim}\,10\%$ contribution to the UHECR flux in the toe region from nearby galaxies with a high star formation rate (SFR), which is favored against isotropy at a confidence level of $4\,\sigma$ above ${\sim}\,40\,$EeV \citep{2018ApJ...853L..29A}. 

Both galaxies hosting active galactic nuclei (AGN) and explosions of young massive stars remain candidates for UHECR correlation studies. The present work aims at establishing a model for the latter type of transient events, which are assumed to be traced by the SFR of their host galaxies. The reconstruction of the 3D distribution of matter on scales ranging from hundreds of Mpc down to few Mpc, including corrections for increasing incompleteness with increasing distance and obscuration by the Galactic Plane, constitutes a prerequisite to cartographying the UHECR sky from the ankle region up to the toe region.

Efforts to establish a 3D mapping of stellar mass within few hundreds of Mpc have recently been led by the gravitational-wave community in the context of galaxy-targeting approaches \citep[e.g.\ GLADE and MANGROVE  catalogs from][respectively]{2018MNRAS.479.2374D, 2020MNRAS.492.4768D}, which aim at identifying the host galaxies of gravitational-wave events or of other transients such as gamma-ray bursts. The community combined spectroscopic surveys with photometric  distances from the 2MASS Photometric Redshift Catalog \citep[2MPZ,][]{2014ApJS..210....9B}, so as to extend the volume of interest with respect to that covered by 2MRS, at the expense of including less accurate distance measurements for galaxies with no spectroscopic distance. 

The present work aims at complementing the GLADE and MANGROVE catalogs established by the gravitational-wave community with cosmic-ladder distance measurements within a few tens of Mpc and with further spectroscopic redshifts for galaxies within the Hubble flow (Sec.~\ref{sec:cata}). The sample established by the GLADE and MANGROVE authors is studied against the mass function established in deep-field observations in Sec.~\ref{sec:mass}, by accounting for increasing incompleteness with increasing distance and proximity to the Galactic plane. The SFR of galaxies within the sample is inferred on a statistical basis from scaling relations with stellar mass depending on galaxy morphology in Sec.~\ref{sec:SFR}. The resulting 1D, 2D, and 3D distributions of stellar mass and SFR are studied in Sec.~\ref{sec:csm_csfr}. The revised sample of galaxies, containing nearly half a million galaxies, is made available for studies against the astroparticle skies in Appendix~\ref{app:revised_sample}. 

The initial mass function (IMF) of \cite{2003PASP..115..763C} is adopted as a reference throughout this paper. Stellar mass and SFR estimates from the literature based on other IMFs are converted, following \cite{2014ARA&A..52..415M}, by multiplying mass and SFR estimates based on the IMF of \cite{1955ApJ...121..161S} by constant factors of 0.61 and 0.63, respectively, and estimates based on the IMF of \cite{2001MNRAS.322..231K} by factors of 0.92 and 0.94, respectively. A flat $\Lambda$CDM cosmology is adopted with $\Omega_M = 0.3$ and $H_0 = 70\,{\rm km\,s}^{-1}\,{\rm Mpc}^{-1}$, corresponding to a critical density $\rho_{\rm c} = \frac{3H_0^2}{8\pi G} = 1.34 \times 10^{11}\,M_\odot\,$Mpc$^{-3}$.

\section{Catalogs} \label{sec:cata}

The distribution of baryonic matter in the local universe and over more than 90\% of the sky has only been mapped by a few surveys. The largest photometric surveys available to date within a dozen and a few hundreds of Mpc are cross matched in this section with the HyperLEDA distance database \citep{2014A&A...570A..13M},\footnote{\url{http://leda.univ-lyon1.fr/}} to establish a controlled sample that is representative of matter distribution on scales ranging from that of the Local Group to distant superclusters.

\subsection{Local Volume}\label{sec:LV}

The most complete full-sky sample of nearby galaxies has been compiled by \cite{2018MNRAS.479.4136K}. The authors have established a census of galaxies within the Local Volume (LV), aimed at being distance limited up to a luminosity distance $d_{\rm L} = 11\,$Mpc. Their sample of 1,029 galaxies compiles tabulated distances, morphologies, stellar masses and SFRs either based on far-ultraviolet \citep[FUV from GALEX, see][]{2017ApJS..230...24B} or dedicated H$_\alpha$ observations undertaken by the authors or from the literature. Twenty eight galaxies observed in the Zone of Avoidance (ZoA) are not included in the sample due to strong Galactic extinction ($A_B>3\,$mag). Luminosity distance estimates are available for all galaxies in the LV sample, with an accuracy ranging from 5 to 10\% for those based on the tip of the red giant branch and up to 25\% for those based on other cosmic-distance ladder methods, such as Cepheids, supernovae, or scaling relations of late- and early-type galaxies \citep[see][and references therein]{2013AJ....145..101K}. All but one galaxy in the LV sample have an assigned morphological type, $T_{\rm LV}$, tabulated on de Vaucouleurs' scale \citep{1991rc3..book.....D}. A particular attention was paid by \cite{2018MNRAS.479.4136K} to the morphological classification of dwarf galaxies, which constitute a significant fraction of LV galaxies, owing to the low-luminosity threshold of this sample. K-band observations from 2MASS at 2.2\,$\mu$m are supplemented with dedicated photometric measurements to estimate stellar masses. At the lack of NIR observations, scaling relations based on B-band magnitude and morphology are further exploited to estimate the K-band luminosity, $L_K$. The stellar mass, $M_{*,\, {\rm LV}}$, is estimated by the authors for 1,022 galaxies assuming a Salpeter IMF, i.e.\ $M_* = 1(M_\odot/L_\odot)\times L_K$, . As noted by \cite{2018MNRAS.479.4136K}, the most up-to-date scaling relations feature a proportionality factor closer to $0.5$ \citep[ranging in $0.5-0.65$ according to][]{2020MNRAS.492.4768D}. The present work assumes, following \cite{2020MNRAS.492.4768D}, $M_{*,\, {\rm LV}} = 0.6 (M_\odot/L_\odot) \times L_K$, which is consistent with a Chabrier IMF. The FUV and H$_\alpha$ SFR estimates of  \cite{2018MNRAS.479.4136K} are corrected for Galactic \citep{2011ApJ...737..103S} and internal extinction, where the latter is estimated from the apparent axial ratio and rotation amplitude of the galaxies \citep[see Sec.~6 in][]{2018MNRAS.479.4136K}. FUV and H$_\alpha$ estimates of the SFR, based on a Salpeter IMF \citep[conversion factors of][respectively]{2011ApJS..192....6L,1998ARA&A..36..189K}  are available for 646 and 467 galaxies out of 1,022 with a stellar-mass estimate.

\subsection{GLADE and MANGROVE}  \label{sec:glade}

The GLADE catalog \citep{2018MNRAS.479.2374D} compiles a full-sky sample of about 800,000 galaxies labeled with a unique identifier, \texttt{idx}. This heterogeneous catalog is built from the GWGC catalog \cite[][including B-band magnitudes]{2011CQGra..28h5016W} and from the 2MASS-XSC survey \cite[][including K-band observations]{2000AJ....119.2498J, 2006AJ....131.1163S}. Luminosity distance estimates are in part extracted from the HyperLEDA database, featuring both cosmic-distance ladder and spectroscopic measurements, and from the 2MPZ catalog, which provides photometric distance estimates. Such photometric redshifts are derived by \cite{2014ApJS..210....9B} with an artificial-neural-network analysis of a cross matched sample from 2MASS-XSC (J, H, K bands at 1.3, 1.7 and 2.2\,$\mu$m), WISE \citep[W1 and W2 bands at 3.4 and 4.6\,$\mu$m, see][]{2010AJ....140.1868W} and SuperCOSMOS \citep[SCOS in the following, with B, R, I bands at 0.45, 0.66 and 0.81\,$\mu$m, see][]{2001MNRAS.326.1295H}. The resolution on individual 2MPZ photometric distances is estimated at 12\%. Spectroscopic estimates from 2MRS are also available for about 45,000 objects out to 400\,Mpc, which provides by itself a complete mapping of stellar mass over more than 90\% of the sky (ZoA excluded) up to limiting K-band magnitudes of 11.75 mag.

As only about two thirds of galaxies within 400\,Mpc were associated with K-band photometry by the GLADE authors,  \cite{2020MNRAS.492.4768D} supplemented GLADE with W1-band magnitude from the AllWISE sample \citep{2010AJ....140.1868W}, resulting in the MANGROVE sample. The association to WISE objects allowed the MANGROVE team to identify strong NIR AGN emission in 3,300 GLADE objects, based on their W1 to W2 flux ratio. \cite{2020MNRAS.492.4768D} derived stellar masses from the W1 band with an uncertainty of $0.1-0.2\,$dex for 743,780 objects out of 800,986 within 400\,Mpc (93\% match efficiency). The stellar-mass estimates account for dust attenuation and are based on a mass-to-light ratio of 0.6, as indicated in Sec.~\ref{sec:LV}.

\begin{deluxetable}{cc}[t!]
\tablecaption{Revised Principal Galaxy Catalog association to the MANGROVE sample\label{tab:mangrove_pgc}}
\tablehead{\colhead{\quad\qquad Glade IDX}\quad\qquad\qquad & \colhead{\quad\qquad PGC}\quad\qquad\qquad }
\startdata
0 & 2789 \\
1 & 46957 \\
3 & 48082 \\
4 & 43495 \\
5 & 1014 \\
\dots & \dots 
\enddata
\tablecomments{671,593 entries. Available online (see Appendix~\ref{app:revised_sample}).}
\end{deluxetable}

\subsection{HyperLEDA}\label{sec:HL}

The HyperLEDA database, queried on 2020-05-10 for the present work, contains 5,377,527 entries with a unique \texttt{pgc} number, an associated HyperLEDA name, equatorial coordinates, and a wealth of information on each of the galaxies. HyperLEDA is constructed as an assembly of the LV sample, the extragalactic distance database \citep{2009AJ....138..323T} and the NASA/IPAC extragalactic database (NED),\footnote{\url{https://ned.ipac.caltech.edu/}} supplemented with relevant works from the literature. Distance estimates are available for more than half of the sample (about 2.9 million entries), as compiled from over 2,400 bibliographic references with substantial contributions from SDSS-III DR12 \citep{2015ApJS..219...12A}, the 2dF final data release \citep{2003astro.ph..6581C} and the 6dF final data release \citep[DR3,][]{2009MNRAS.399..683J}.\footnote{\url{http://leda.univ-lyon1.fr/a110}} The HyperLEDA distance estimates are mostly based on spectroscopic measurements, but also include constraints from the cosmic-distance ladder for 4,237 entries. Although B-band and K-band magnitudes are available for 84\% and 33\% of the objects, the HyperLEDA assembly is not aimed at being flux limited, so that a selection procedure is needed to ensure uniform coverage over the sky. The HyperLEDA database contains many pieces of information on galaxies in the local universe, most of which is not relevant to the present work. Noteworthy though is the morphological type tabulated on de Vaucouleurs' scale, $T_{\rm HL}$, which is available for 10\% of the galaxies and is exploited in Sec.~\ref{sec:SFR}.

\begin{deluxetable}{cc}[t!]
\tablecaption{Duplicate entries in the MANGROVE sample\label{tab:mangrove_dupl}}
\tablehead{\colhead{Glade IDX} & \colhead{Glade IDX duplicates}}
\startdata
3623 & 381647 \\
4900 & 266006 \\
9755 & 314233 \\
12714 & 282396 \\
13246 & 282405 \\
\dots & \dots
\enddata
\tablecomments{361 entries. Available online (see Appendix~\ref{app:revised_sample}).}
\end{deluxetable}

\subsection{Pairing and cleaning} \label{sec:pairing_cleaning}

The information in the GLADE and MANGROVE catalogs inherited from HyperLEDA is limited: while two fields are expected to provide the associated name and \texttt{pgc} number of the galaxy, these quantities are missing for 187,579 and 748,809 entries, respectively, out of 800,986. To circumvent this limitation, a systematic check of the \texttt{pgc}, \texttt{GWGC\_name}, \texttt{HyperLEDA\_name}, \texttt{2MASS\_name} and \texttt{SDSS-DR12\_name} entries of MANGROVE objects with an estimated mass is performed. Each numerical entry in the \texttt{pgc} and \texttt{HyperLEDA\_name} fields consistent with being an integer is stored as a putative \texttt{pgc} number. The entries of the five above-mentioned fields consistent with being strings are then searched for in HyperLEDA, through an automated search and case-by-case queries to HyperLEDA, as well as to NED and the SIMBAD Astronomical Database\footnote{\url{http://simbad.u-strasbg.fr/simbad/}} for ambiguous entries. This procedure results in associating a unique \texttt{pgc} identifier to about 90\% of the MANGROVE objects with stellar mass, as shown in Table~\ref{tab:mangrove_pgc}. The remaining 10\% consist of 2MASS objects, with empty \texttt{pgc}, \texttt{GWGC\_name}, \texttt{HyperLEDA\_name} and \texttt{SDSS-DR12\_name} fields. These 2MASS objects were not present in the HyperLEDA database at the time of the query, as confirmed by the maintainers.\footnote{D. Makarov, private communication.}

Redundant entries are identified through their \texttt{2MASS\_name} and \texttt{pgc} numbers, whenever available. Each of the duplicates listed in Table~\ref{tab:mangrove_dupl} is confirmed to be a twin through dedicated HyperLEDA, NED and SIMBAD searches. Duplicate entries are mostly formed of a HyperLEDA object and a 2MASS object, which are not associated in the GLADE sample, in addition to a small fraction of doubled HyperLEDA entries with two different identifiers (older and newer \texttt{pgc}).  As a result of the association work performed by the GLADE and MANGROVE teams, duplicates represent only ${\sim}\,5$ out of 10,000 entries. Only the brightest associated WISE source of each pair is kept, as listed in the first column of Table~\ref{tab:mangrove_dupl}.

\begin{deluxetable}{cc}[t!]
\tablecaption{Principal Galaxy Catalog association to the Local Volume sample\label{tab:lv_pgc}}
\tablehead{\colhead{\quad Local Volume Name}\quad\qquad &\colhead{\quad PGC} \quad\qquad}
\startdata
A0554+07 & 138836 \\
A0952+69 & 28759 \\
AF7448\_001 & 5067080 \\
AGC112454 & 1456468 \\
AGC112503 & 3339344 \\
\dots & \dots
\enddata
\tablecomments{1,029 entries. Available online (see Appendix~\ref{app:revised_sample}).}
\end{deluxetable}

The \texttt{pgc} field enables a straightforward pairing of galaxies from multiple samples. Luminosity distances are revised as follows for objects present in multiple samples: LV distance estimates, checked by their authors on a case-by-case basis, are preferred, followed by HyperLEDA distance estimates (\texttt{modbest} field). Distances tabulated in GLADE are used at the lack of other estimates. The GLADE authors provide a distance-flag field, \texttt{flag\_d}, distinguishing spectroscopic estimates (\texttt{flag\_d=2, 3}) from 2MPZ photometric estimates (\texttt{flag\_d=1}). Following the revision performed in the present work, the distance-flag field is supplemented with \texttt{flag\_d=4} for LV distances, \texttt{flag\_d=5} for purely spectroscopic estimates and \texttt{flag\_d=6} for cosmic-ladder estimates from HyperLEDA (non-empty \texttt{mod0} field).

\begin{deluxetable}{cc}[t]
\tablecaption{Entries excluded from the MANGROVE sample\label{tab:excluded}}
\tablehead{\colhead{Glade IDX} & \colhead{Comment}}
\startdata
213 & Dubious GLADE - WISE association \\
481 & gam AGN \\
487 & gam AGN \\
539 & BLL \\
594 & Radio Galaxy \\
\dots & \dots
\enddata
\tablecomments{1,387 entries. Available online (see Appendix~\ref{app:revised_sample}).}
\end{deluxetable}

\begin{figure*}[t]
\epsscale{1.2}
\plotone{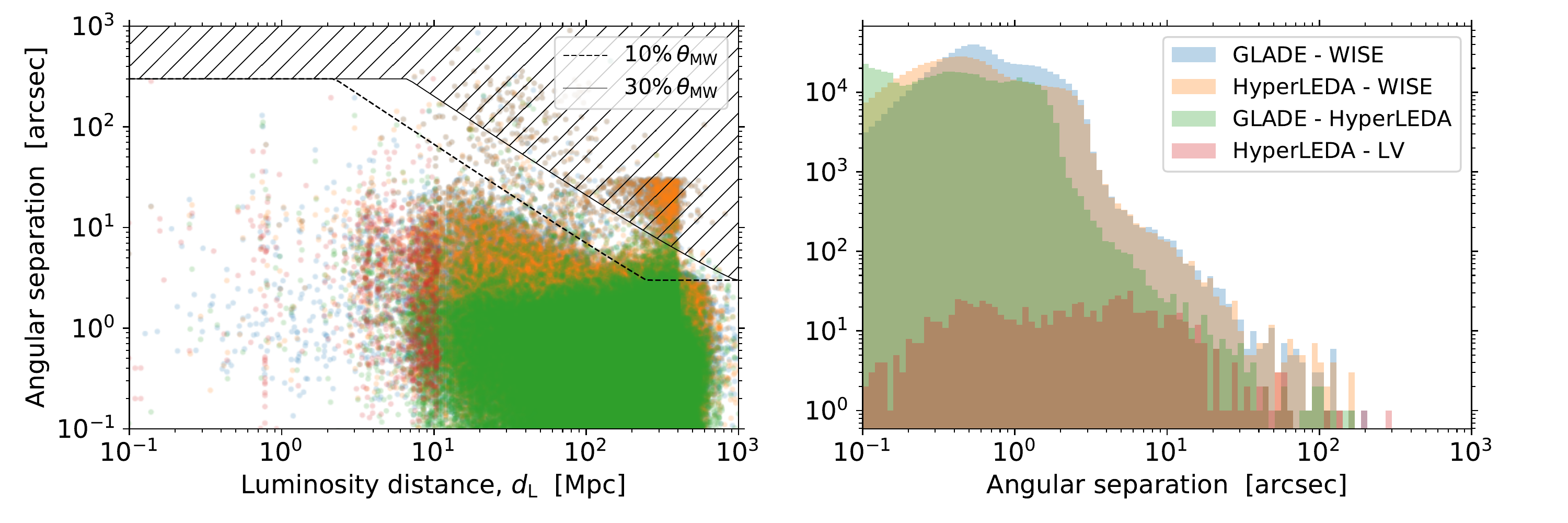}
\caption{\textit{Left}:  The angular separation between paired entries from HyperLEDA, GLADE and WISE, and between paired entries from the LV and HyperLEDA samples as a function of revised luminosity distance. The dashed and solid black lines mark an angular extent of 10 and 30\%, respectively, of the diameter of the Milky Way placed at a given distance, limited to 5\,arcmin at small distances and to 3\,arcsec at large distances. The hashed region, encompassing a cluster of entries separated by ${\sim}\,20\,$arcsec at revised distances beyond 100\,Mpc, is excluded in the present study. \textit{Right}: The distributions of the angular separation between paired entries retained in the present study.\label{fig:angular_distance_cut}}
\end{figure*}

A study of possible misassociation is performed by investigating the angular separation between the WISE, GLADE and HyperLEDA paired entries, as shown in Fig.~\ref{fig:angular_distance_cut}. For reference, the angular separation between LV galaxies and associated HyperLEDA entry (see Table~\ref{tab:lv_pgc}) is also displayed. A vast majority of the objects are associated across the four catalogs within a few arcsec. A sharp cutoff in the GLADE-WISE angular separation is observed at about 20\,arcsec, which corresponds to the maximum distance imposed by the MANGROVE authors. In the latter study, the authors re-associated GLADE entries to galaxies from the AllWISE sample, at the lack of a 2MPZ association field in the GLADE catalog. The spatial cross match was performed within an angular radius corresponding to 5\% of the angular diameter of the Milky Way (32.4\,kpc), limited to 20\,arcsec at low distances and 3\,arcsec at large distances. Single-entry associations within the W1 point-spread function, of extent 6.1\,arcsec, were also accepted. The revision of distances performed in the present work reveals a small fraction (1,254 entries) of dubious associations, particularly marked by a cluster of entries at revised distances larger 100\,Mpc in Fig.~\ref{fig:angular_distance_cut}, left. These entries are excluded by a cut at 30\% of the angular diameter of the Milky Way. A maximum separation of 5\,arcmin is adopted for nearby galaxies, corresponding to the maximum separation between HyperLEDA and LV entries (see Fig.~\ref{fig:angular_distance_cut}, right). Entries in the buffer area between 10 and 30\% of the angular diameter of the Milky Way are considered as tentative associations (\texttt{flag\_asso=2}), while entries below the buffer area correspond to more robust associations (\texttt{flag\_asso=1}). The results presented in the following sections are unaffected by the inclusion or exclusion of the tentatively associated entries, which are kept in the revised MANGROVE sample. 

\subsection{Revised MANGROVE sample} \label{sec:revised_sample}

The galaxy sample studied in the present work consists of the assembly of galaxies with a stellar-mass estimate extracted from the LV and MANGROVE samples. Stellar masses tabulated in the MANGROVE sample, based on W1-band observations (\texttt{flag\_M*=1}), are updated according to the revised luminosity distance estimates, $d_{\rm L}$, while stellar masses from the LV sample, based on K-band observations (\texttt{flag\_M*=0}), are left unchanged by the distance revision.

The 1,101 most-massive galaxies with $\log_{10}M/M_\odot > 11.5$ are investigated on a case-by-case basis through SIMBAD queries. Most are confirmed to be bright galaxies dominated by stellar emission in the NIR. Four stars are identified and are removed from the sample. Jetted AGN, in the form of radio galaxies or blazars \citep[in particular BL-Lac type blazars which dominate over flat-spectrum radio quasars in the local universe, see e.g.][]{2020ApJ...892..105A}, represent less than 10\% of the entries at $\log_{10}M/M_\odot > 11.5$. They are also excluded from the sample, as contamination from the jet emission could bias the host-galaxy stellar-mass estimate. All less-massive objects classified as BL Lacs in HyperLEDA are further excluded. Seyfert galaxies and quasi-stellar objects, often hosted by spiral galaxies which can show significant star-forming activity, are kept in the sample. The list of excluded objects, including the paired entries with large angular separations identified in Sec.~\ref{sec:pairing_cleaning}, is provided in Table~\ref{tab:excluded}. The comment field provides the principal type of the object in SIMBAD, corresponding to a jetted AGN or a star, or a note on the outliers at large angular separations.

\begin{figure}[t!]
\epsscale{1.2}
\plotone{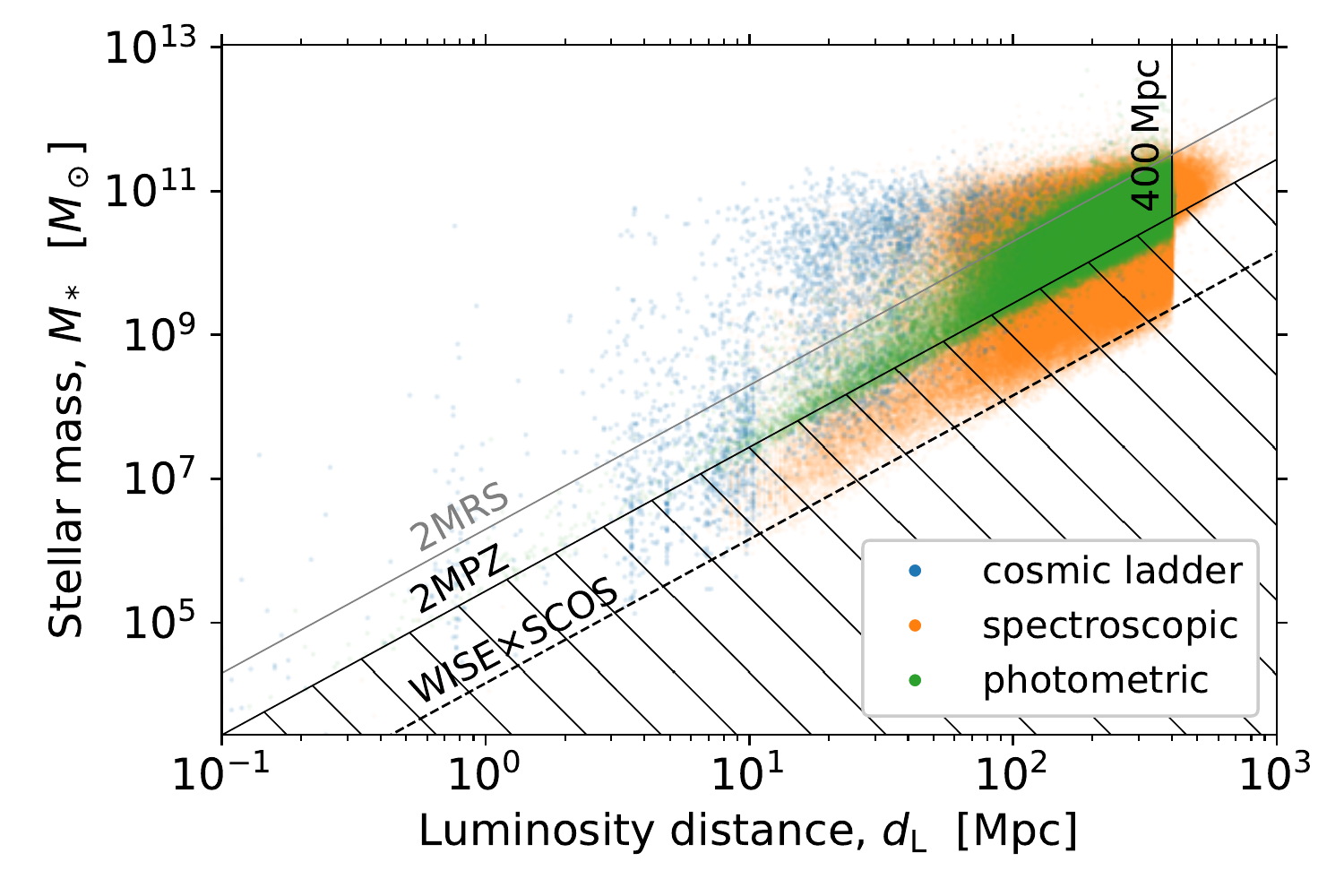}
\caption{The stellar masses of galaxies versus revised luminosity distances from cosmic-ladder (blue), spectroscopic (orange) and photometric (green) estimates. The flux limits of 2MRS, 2MPZ and WISE$\times$SCOS are indicated as solid gray, solid black and dashed black lines, respectively. For reference, the 400\,Mpc cut placed by the MANGROVE authors is indicated as a vertical solid line. The distance of galaxies farther away are derived from spectroscopic measurements, superseding the initial  2MPZ photometric estimate. The hashed region is excluded in the present study.\label{fig:mass_cut}}
\end{figure}

Stellar masses from the revised MANGROVE sample are shown as a function of luminosity distances in Fig.~\ref{fig:mass_cut}. Cosmic-ladder distance estimates (\texttt{flag\_d=4,6}) predominate at distances lower than 50\,Mpc, followed at larger distances by spectroscopic estimates for the brightest objects above the 2MRS sensitivity at ${\rm K} = 11.75\,$mag. Photometric estimates constitute the main source of distance information for fainter objects down to the 2MPZ sensitivity limit at ${\rm K} = 13.9\,$mag. The sensitivity limits of the 2MRS and 2MPZ samples at 10\,Mpc are estimated to correspond to $\log_{10}M_{\rm 2MRS}(10\,\mathrm{Mpc})/M_\odot = 8.40$ and $\log_{10}M_{\rm 2MPZ}(10\,\mathrm{Mpc})/M_\odot = 7.54$, respectively. The completeness in flux of the sample cannot be ensured below the 2MPZ sensitivity. Supplementing the GLADE/MANGROVE sample with photometric distance estimates from WISE$\times$SCOS \citep{2016ApJS..225....5B} would be needed to exploit the full potential of WISE up to a W1 magnitude of 17, which is estimated to correspond to $\log_{10}M_*/M_\odot = 6.26$ at 10\,Mpc. Only objects above the 2MPZ sensitivity limit are considered in the remainder of the present work, with no a priori limitation in distance. Nonetheless, the distance cut at 400\,Mpc imposed by the MANGROVE authors, combined with the 12\% distance resolution from photometric estimates, suggests that deviations from a flux-limited sample could be expected starting at around 350\,Mpc.\footnote{The impact of distance resolution effects is illustrated by the orange tail in the top-right corner of Fig.~\ref{fig:mass_cut}: these are galaxies with a photometric estimate in GLADE and a revised spectroscopic distance in the present work.}

The sample studied in the present work consists of 489,008 galaxies above the 2MPZ sensitivity limit, 394,209 of which are located at distances lower than 350\,Mpc. Among 394,209 galaxies within 350\,Mpc, 3,198 distances are based on the cosmic-distance ladder, 190,290 are spectroscopic (only 137 of these are from GLADE and not from HyperLEDA) and 200,721 are based on photometric redshifts. These numbers can be compared to 48,825 and 304,147 spectroscopic and photometric estimates, respectively, out of 352,972 objects above the 2MPZ sensitivity limit within 350\,Mpc in the initial MANGROVE catalog.  The revision of distances performed in Sec.~\ref{sec:pairing_cleaning} thus not only enables to increase the fraction of objects with a spectroscopic distance by a factor of four but also to include in the revised sample about ${\sim}\,50,000$ objects that otherwise would have been assumed to lie beyond 350\,Mpc based on photometric distance estimates.

\newpage
\section{Stellar Mass} \label{sec:mass}

The sample selected in Sec.~\ref{sec:revised_sample} consists essentially of galaxies from the 2MASS-XSC and AllWISE catalogs and is limited in flux by the sensitivity of the former survey. \cite{2011MNRAS.416.2840L} studied a subset of the sample scrutinized in the present work, limited in the K-band to 11.5\,mag or 12.5\,mag depending on the region of the sky. Exploiting spectroscopic distance estimates from 2MRS, SDSS \citep{2015ApJS..219...12A} and 6dF \citep{2009MNRAS.399..683J}, the authors studied the mass function of galaxies in their sample and proposed weights to account for incompleteness outside the ZoA, defined as a band of $5^\circ$ around the Galactic plane, with a wider extent of $10^\circ$ at Galactic latitudes $|l|<30^\circ$. Employing galaxy cloning to fill the ZoA, \cite{2011MNRAS.416.2840L} achieved a tomographic mapping of stellar mass up to $285\,(h/0.7)^{-1}\,$Mpc in the regions with the deepest spectroscopic coverage. Following the spirit of \cite{2011MNRAS.416.2840L}, weights are proposed in the present work to account for variations of completeness near the ZoA and as a function of distance, exploiting for the latter the mass function of galaxies derived from the Galaxy And Mass Assembly \citep[GAMA,][]{2017MNRAS.470..283W}. Deep-field observations of the GAMA region covering ${\sim}\,1\%$ of the sky up to $z=0.1$ enabled \cite{2017MNRAS.470..283W} to model the mass function with a double Schechter function down to $10^{6.5}\,M_{\odot}$, which corresponds to the sensitivity limit of 2MPZ at about 3\,Mpc. As in \cite{2011MNRAS.416.2840L}, galaxy cloning is employed in the ZoA in order to minimize biases induced by a shallow coverage over a wide band of the sky \citep[see also][]{2011ApJ...741...31B} and thus enable large-scale correlation searches with astroparticles. The alternative approach, namely masking the Galactic plane, is inappropriate for astroparticles whose arrival directions are affected by large uncertainties or by deflections in intervening magnetic fields.

\begin{figure*}[ht!]
\epsscale{1.2}
\plotone{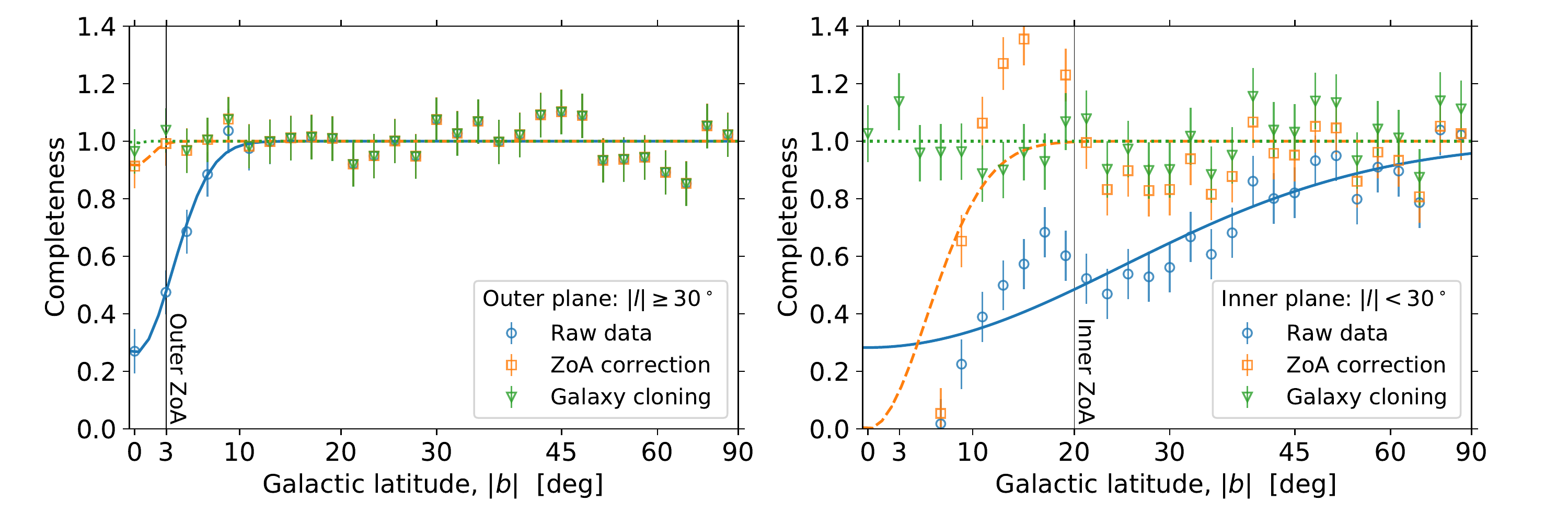}
\caption{The relative completeness as a function of Galactic latitude at outer ($|l|\geq30^\circ$, \textit{left}) and inner ($|l|<30^\circ$, \textit{right}) Galactic longitudes. The relative completeness is estimated with galaxies at luminosity distances ranging in $100-300\,$Mpc and is shown as blue circles prior to any correction, as orange squares after ZoA correction and as green triangles after further galaxy cloning in the regions marked by the vertical black lines.\label{fig:completeness}}
\end{figure*}

\subsection{Treatment of the Zone of Avoidance}\label{sec:ZoA}

The galaxy counts of the revised MANGROVE sample determined in 30 equal-area bins in Galactic latitude, $b$, are shown in Fig.~\ref{fig:completeness}, for galaxies at luminosity distances ranging in $100-300\,$Mpc. As shown in Sec.~\ref{sec:csm_csfr} and \cite{2018AN....339..615K}, the distribution of stellar mass is expected to be close to isotropic on large angular scales beyond 100\,Mpc. An upper bound at 300\,Mpc for the distance range is used to avoid any possible bias induced by the distance limit imposed in the MANGROVE sample. Uncertainties on number counts displayed in Fig.~\ref{fig:completeness} account for both the Poisson statistics in each bin and for cosmic variance, estimated from bin-to-bin variations at Galactic latitudes $|b|>45^\circ$, where ZoA corrections are assumed to be small. The drop in galaxy counts close to the ZoA (see ``Raw data'' in Fig.~\ref{fig:completeness}) is empirically modeled by a Gaussian function of $\sin b$, with correction factors

\begin{equation}
c_b(b)= 1-c_0 \times \exp\Big(-\frac{(\sin b)^2}{2(\sin b_{\rm ZoA})^2}\Big). 
\label{eq:galcorr}
\end{equation}

The correction is estimated to go up to $c_{0,\, {\rm inner}} = 72 \pm 6\,\%$ and $c_{0,\, {\rm outer}} = 74 \pm 6\,\%$ in the inner ($|l|<30^\circ$) and outer ($|l|\geq30^\circ$) Galactic plane regions, with Gaussian extents of $b_{\rm ZoA,\, inner}=25^\circ \pm 5^\circ$ and $b_{\rm ZoA,\, inner}=3.6^\circ \pm 0.3^\circ$, respectively. Weighting each galaxy count by the inverse of $c_b(b)$ results in a Galactic latitude distribution compatible with isotropy within cosmic variance in the outer plane, but the correction is not sufficient to account for the count drop down to zero near the Galactic center (see Fig.~\ref{fig:completeness}, orange squares labeled as ``ZoA correction''). 

As in  \cite{2011MNRAS.416.2840L}, galaxies within the ZoA are removed from the sample and replaced with mirrored counterparts from equal-area regions above and below the ZoA, so that each cloned galaxy conserves the properties of its true counterpart outside the ZoA, including the correction factor in Eq.~\ref{eq:galcorr}, except for the Galactic latitude

\begin{equation}
\sin b_{\rm clone} = \sin b_{\rm ZoA} - (\sin b - \sin b_{\rm ZoA}),
\end{equation}
where $\sin b \in [\sin b_{\rm ZoA}, 2 \sin b_{\rm ZoA}]$. In this way, galaxies right above the ZoA are cloned right below the ZoA boundary and the Galactic plane is effectively filled with galaxies at $\sin b = 2 \sin b_{\rm ZoA}$. Galaxy cloning preserves the large scale properties of the 3D distribution of galaxies at the expense of replacing true small-scale anisotropies close to the Galactic plane by those in the band $[\sin b_{\rm ZoA}, 2 \sin b_{\rm ZoA}]$. 

The regions filled with galaxy clones are defined, as shown in Fig.~\ref{fig:completeness}, as those with $|b|<20^\circ$ and $|b|<3^\circ$ in the inner and outer plane, respectively. The cloning algorithm results in 19,274 galaxies and 40,338 clones being removed from and added to the sample, respectively. It is thus inferred that about half of the galaxies in the ZoA are missing from the initial sample. The above-defined ZoA covers 10\% of the sky, so that ${\sim}\,5\%$ of the galaxies above the 2MPZ sensitivity are missed because of their location in the ZoA. The correction factors and cloning procedure are not applied at $d_{\rm L} < 11\,{\rm Mpc}$, as bright galaxies close to the Galactic plane in this distance range could result in anisotropies relevant to full-sky studies.\footnote{Section~\ref{sec:LV} indicates that objects with high Galactic extinction are not included in the LV sample of \cite{2018MNRAS.479.4136K}. It was checked that the merging with the GLADE-MANGROVE catalog makes up for the deficit in the ZoA.} The final galaxy sample consists of 510,072 galaxies, including 40,338 clones, whose latitude distribution beyond 100\,Mpc is consistent, within cosmic variance, with isotropy on large angular scales (see Fig.~\ref{fig:completeness}, green triangles labeled as ``Galaxy cloning''). Given the adopted ZoA limits at $|b|<20^\circ$ and $|b|<3^\circ$ in the inner and outer plane, respectively, the ZoA correction factors, $c_b(b)$, of galaxies and clones kept in the sample range in $0.48-1$, so that the count weights of individual galaxies resulting from the ZoA correction, $\{1/c_b(b_i)\}_i$, do not exceed a factor on the order of two.

\subsection{Completeness as a function of distance}

\begin{figure*}[ht!]
\epsscale{1.2}
\plotone{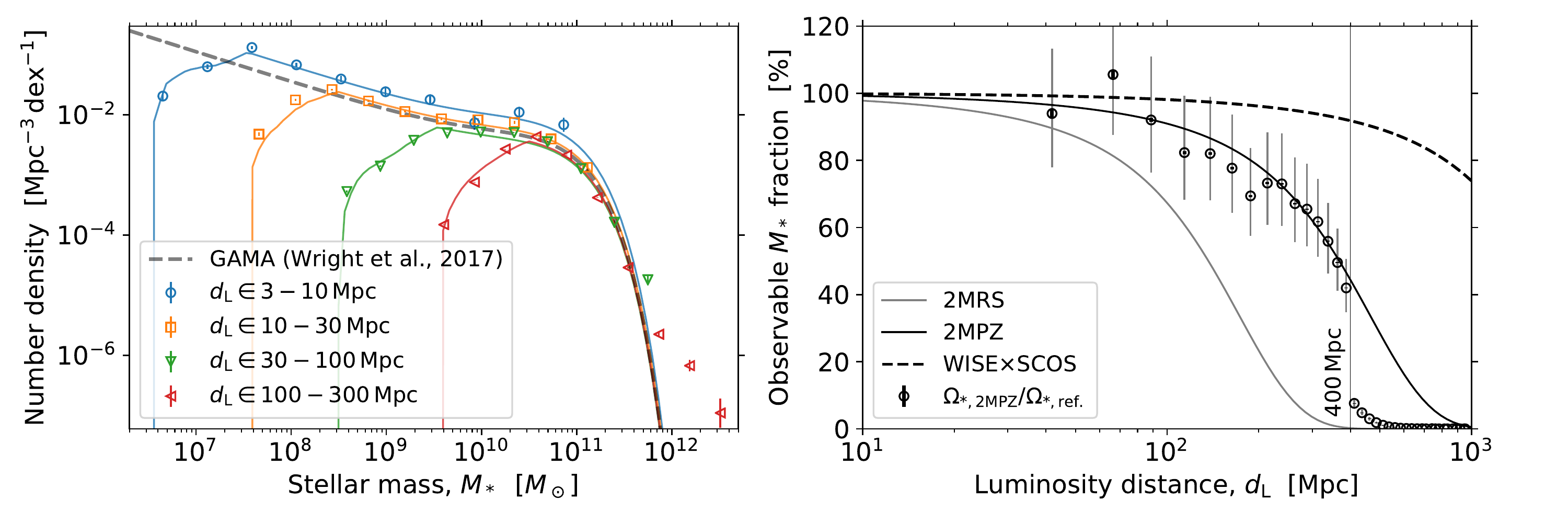}
\caption{\textit{Left}: The stellar-mass functions over the entire sky, after ZoA correction and galaxy cloning, for luminosity-distance ranges labeled in the figure. The stellar-mass function estimated in \cite{2017MNRAS.470..283W} is shown for reference as a gray dashed line. Colored lines show the scaled mass functions accounting for the 2MPZ sensitivity limit. \textit{Right}: The fraction of observable stellar mass as a function of luminosity distance above the sensitivity threshold of 2MRS (solid gray line), 2MPZ (solid black line) and WISE$\times$SCOS (dashed black line). The 400\,Mpc cut placed by the MANGROVE authors is indicated as a vertical solid line. The black points display the observed stellar-mass density, $\Omega_{*,\,{\rm 2MPZ}}$, in 25\,Mpc-thick layers beyond the Virgo cluster (${>}\,25\,$Mpc), normalized  to a reference value, $\Omega_{*,\,{\rm ref.}}$, from the GAMA/G10-COSMOS/3D-HST survey of galaxies at $z\in[0.02,0.08]$ \citep{2018MNRAS.475.2891D}. Points beyond 400\,Mpc correspond to galaxies with a photometric distance in the MANGROVE sample superseded by a spectroscopic estimate in the present work. The uncertainties on the density ratios are dominated by correlated uncertainties on $\Omega_{*,\,{\rm ref.}}$ (thin gray lines) rather than by uncorrelated uncertainties on galaxy counts (thick gray lines within markers). \label{fig:m_fun}}
\end{figure*}

The stellar-mass functions of galaxies within four distance ranges, accounting for ZoA count weights and including galaxy clones in the ZoA, are shown in Fig.~\ref{fig:m_fun}, left. Beyond 10\,Mpc and above the sensitivity threshold for each distance range, the reconstructed mass functions agree at the $20-30\%$ level with that derived from GAMA deep-field observations, which is represented as a dashed gray line in Fig.~\ref{fig:m_fun}, left \citep[bolometric estimate in][]{2017MNRAS.470..283W}. For comparison, a $\sim 16\%$ statistical uncertainty is estimated on the GAMA mass function. The higher normalization of observed counts at distances lower than 10\,Mpc is expected from the overdensity of matter in the LV \citep[see Sec.~\ref{sec:csm_csfr} and][]{2018AN....339..615K}. 

At the low-mass end of the luminosity function, the sensitivity threshold is accounted for in Fig.~\ref{fig:m_fun} by estimating the fraction of observable galaxies at a given stellar mass, provided the observed distance distribution in each distance range. The resulting model for the observed stellar-mass functions are scaled so that the total expected number of galaxies in each distance range is equal to the weighted sum of observed galaxy counts, $\sum_i 1/c_b(b_i)$. 

The high-mass tail of the full-sky sample above $10^{11.5}\,M_{\odot}$ is not expected from the best-fit GAMA model, which is unable to sample the lowest number densities at the very high-mass end due to a smaller sky coverage. Such a discrepancy cannot be explained by resolution effects induced by photometric estimates: a 12\% uncertainty in distance results in a stellar-mass uncertainty of 0.1\,dex. Similarly, a cross-match with the GAMA sample from \cite{2020MNRAS.498.5581B},\footnote{Private communication from S.~Bellstedt and S.~Driver on behalf of the GAMA team, see also upcoming GAMA Data Release 4.} resulting in the selection of about 700 galaxies with spectroscopic distances, stellar masses above the 2MPZ sensitivity limit and robust SFR estimates (uncertainty smaller than 1\,dex), indicates consistency in stellar-mass measurements with a dispersion better than 0.15\,dex and a bias smaller than 0.05\,dex. As indicated in Sec.~\ref{sec:revised_sample}, known jetted AGN have been removed from the sample under study. Non-jetted active galaxies classified as AGNs, quasars, LINERs and Seyfert galaxies represent 7\% of the galaxies with stellar mass above $10^{11.5}\,M_\odot$, so that AGN contamination is unlikely to explain the presence of such a tail. About 20\% of the galaxies above $10^{11.5}\,M_{\odot}$ are labelled in SIMBAD as brightest galaxy in a cluster (BCG) and about 8\% are instead labelled as belonging to a cluster, group or pair of galaxies. While source confusion or blending of galaxies cannot be excluded as possible explanations for the difference between the GAMA model and the MANGROVE high-mass tail, the present sample could offer an opportunity for targeted investigations of the thousand  galaxies identified in the present work as possibly lying above  $10^{11.5}\,M_{\odot}$, e.g.\ in continuity with the investigations carried out by \cite{2013MNRAS.436..697B, 2017MNRAS.467.2217B}. Such studies would help determine whether the high-mass tail displayed in Fig.~\ref{fig:m_fun} is a new component of the galaxy stellar-mass function or a mere artifact of the GLADE/MANGROVE association procedure.

The overall match of the full-sky mass function with that derived from deep-field observations below $10^{11.5}\,M_{\odot}$ suggests that the sample under-study is consistent with being flux limited. The completeness in total stellar mass can then be estimated from the integral of the GAMA mass function (double Schechter function) above the 2MPZ sensitivity limit:

\begin{equation}
\label{eq:cM}
c_{M}(d) = \frac{\sum_{i=1}^{2} \phi_i \times \Gamma\big(2+\alpha_i, \frac{M_{\rm 2MPZ}(d)}{M_{\rm cut}}\big)}{\sum_{i=1}^{2} \phi_i \times \Gamma\big(2+\alpha_i)},
\end{equation}
where $\Gamma(a)$ is the gamma function, $\Gamma(a,x)$ is the upper incomplete gamma function and $M_{\rm 2MPZ}(d) = M_{\rm 2MPZ}(10\,\mathrm{Mpc}) \times (d/10\,\mathrm{Mpc})^2$. The other parameters are those of the GAMA Schechter functions: $(\phi_1, \phi_2) = (2.93, 0.63) \times 10^{-3}\,{\rm Mpc}^{-3}$, $(\alpha_1, \alpha_2) = (-0.62, -1.5)$ and $\log_{10} M_{\rm cut}/M_\odot = 10.78$ \citep{2017MNRAS.470..283W}.  

The completeness inferred from Eq.~\ref{eq:cM} is compared in Fig.~\ref{fig:m_fun}, right, to the stellar-mass density estimated in 25-Mpc thick layers in luminosity distance, normalized to that estimated by \cite{2018MNRAS.475.2891D} from galaxies at $z\in[0.02,0.08]$ in the GAMA/G10-COSMOS/3D-HST fields. Good agreement is found up to 400\,Mpc within the ${\sim}\,20\%$ uncertainty estimated from deep-field observations. As for ZoA correction factors, galaxy counts are weighted by the inverse of $c_{M}(d)$ in the remainder of the present work, with $c_{M}(d)$ ranging in $[0.53,1]$ for galaxies with luminosity distance lower than 350\,Mpc. The final  weights accounting for stellar-mass completeness, $c_{N, m}(d,b) = c_M(d) \times c_b(b)$, range in $[0.26,1]$ in $350\,{\rm Mpc}\times 4\pi$, so that the count weights of individual galaxies, $\{1/c_{N, m}(d_i,b_i)\}_i$,  do not exceed a factor of four. 

Figure~\ref{fig:m_fun}, right, also displays completeness estimates as a function of luminosity distance above the 2MRS and WISE$\times$SCOS sensitivity limits. 2MPZ photometric distance estimates enable an extension of the probed distance range at similar mass completeness by a factor of ${\sim}\,2.8$ with respect to 2MRS. A further increase by a factor of ${\sim}\,4$ could be expected from a full-sky sample exploiting WISE$\times$SCOS.

\section{Star formation rate} \label{sec:SFR}

As discussed in Sec.~\ref{sec:intro}, the present work aims at establishing a full-sky model for transient astroparticle events traced by star formation. Multiple tracers of star formation are discussed in the literature, ranging from radio and X-ray emission from electrons freshly accelerated in stellar explosions,  far-infrared emission from dust heated by starlight, FUV starlight escaping from the galaxy, to H$_\alpha$ emission from gas ionized by neighbouring massive stars  \citep[see][for reviews]{2013seg..book..419C, 2014ARA&A..52..415M}. Both FUV and H$_\alpha$ tracers of SFR are available in the LV sample. The FUV band traces emission from OB stars on timescales of $100-300$\,Myrs. The H$_\alpha$ flux traces emission from young massive stars ($>20\,M_\odot$) on timescales of $5-10$\,Myrs. Aiming at inferring the most recent activity in the star-formation history of the galaxies, the H$_\alpha$ tracer is adopted as a reference.

\subsection{Intercalibration of SFR tracers}
\label{sec:sfr_calib}

SFR tracers are intercalibrated using the sub-sample of 314 LV galaxies above the 2MPZ sensitivity limit with stellar mass, morphological type, H$_\alpha$ and FUV SFRs. The intercalibration is performed by testing the compatibility of the data with a linear dependence of $s_\alpha = \log_{10}{\rm SFR(H_\alpha)}/{\rm SFR}_0$ on $s_{\rm FUV} = \log_{10}{\rm SFR(FUV)}/{\rm SFR}_0$, where SFR$_0 = 1\,M_\odot\,{\rm yr}^{-1}$ is taken as a reference. With no SFR uncertainty at hand, the dispersion around the relation $s_\alpha = a + b\times s_{\rm FUV}$ is modeled as a normal distribution of free width, $\sigma_{\rm SFR}$ in dex. The best fit-parameters, $a(T)$ and $b(T)$ which are assumed to depend upon morphology, $T$, are determined by maximizing the likelihood

\begin{equation}
\mathcal{L} = \prod_{i=1}^{314} \mathcal{N}\Big(s_{\alpha,\,i}\, \Bigm|\, a(T_i) + b(T_i)\times s_{{\rm FUV},\,i}, \sigma_{\rm SFR}^2 \Big),
\end{equation}
where the index $i$ runs over the 314 galaxies and where $\mathcal{N}(x\,|\,\mu,\sigma^2)$ is the standard normal function. Maximizing $\mathcal{L}$ is equivalent to minimizing the deviance

\begin{equation}
D = \sum_{i=1}^{314}\Bigg[  
\frac
{\big(s_{\alpha,\, i} - a(T_i) - b(T_i) s_{{\rm FUV},\, i} \big)^2}
{\sigma_{\rm SFR}^2} + 2\ln{\sigma_{\rm SFR}} 
\Bigg],
\end{equation}
where the constant terms are left out. The first term in the sum is a $\chi^2$ element, while the second term results from the normalization of the likelihood function, with free $\sigma_{\rm SFR}$. The quality of the fit is inferred from a Kolmogorov-Smirnov (KS) test for normality of the residuals, which are defined as the square root of each $\chi^2$ element.

\begin{figure*}[ht!]
\epsscale{1.2}
\plotone{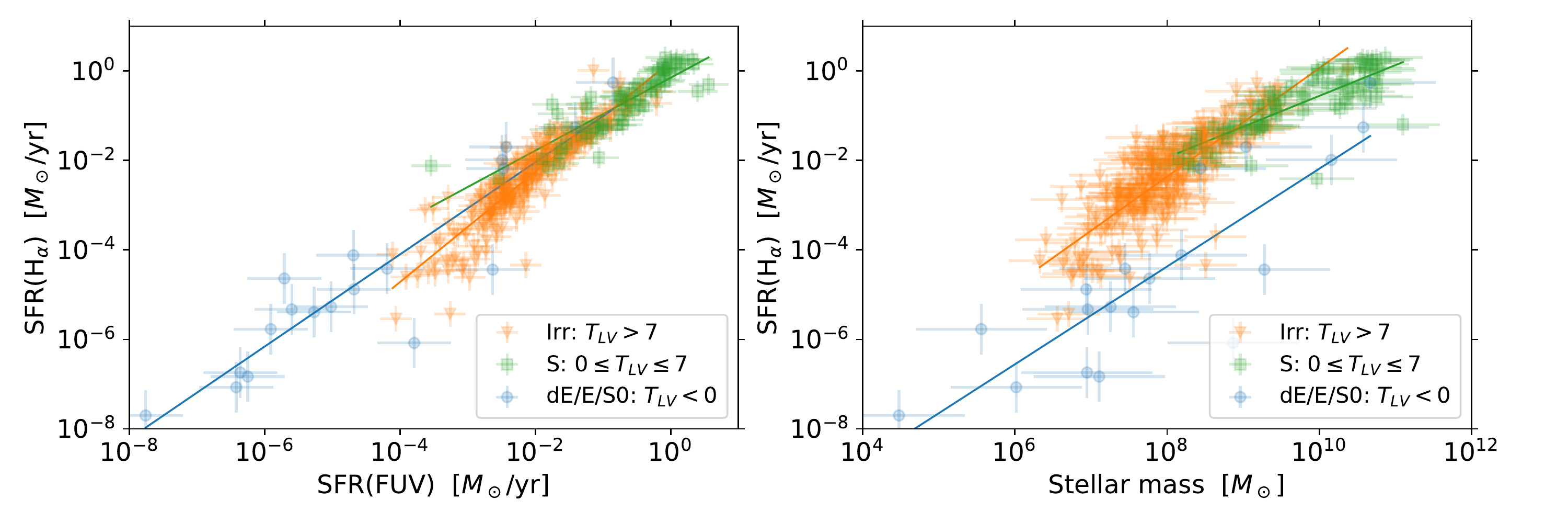}
\caption{The scaling relations of SFR tracers for the three morphological classes, exploiting the ${\rm H}_{\alpha} - {\rm FUV}$ (\textit{left}) and ${\rm H}_{\alpha} - M_*$ (\textit{right}) observations. Elliptical (dE/E/S0), spiral (S) and irregular (Irr) galaxies are shown as blue circles, green squares and orange triangles, respectively. Uncertainties on the SFR tracers are inferred from the best-fit dispersion around the linear relations, as discussed in the text.\label{fig:intercal}}
\end{figure*}

The best-quality results, with a KS-test p-value of 5\%, are obtained by dividing the 314 galaxies into three morphology ranges: $T_{\rm LV}<0$, $0 \leq T_{\rm LV} \leq 7$ and $T_{\rm LV}>7$, labeled as ``elliptical'',\footnote{The morphological class labeled as ``elliptical'' includes dwarf elliptical, elliptical and lenticular galaxies.} ``spiral'' and ``irregular'' galaxies for convenience. The best-fit H$_\alpha$-FUV scaling relations are

\begin{align}
    s_\alpha &= 0.02 + (1.03 \pm 0.09)\times s_{\rm FUV} &\text{ for } T_{\rm LV}<0, \nonumber\\
    &= -0.15 + (0.81 \pm 0.06)\times s_{\rm FUV} &\text{ for } 0\leq T_{\rm LV} \leq 7, \nonumber\\
    &= 0.21 + (1.23 \pm 0.03)\times s_{\rm FUV} &\text{ for } T_{\rm LV}>7,
\end{align}
with dispersion around the best-fit relation of $\sigma_{\rm SFR} = 0.80, 0.34, 0.41\,$dex for elliptical, spiral and irregular galaxies, respectively. Assigning half of the variance to $s_\alpha$ and $b(T_{\rm LV}) s_{\rm FUV}$ results in an effective uncertainty on the H$_\alpha$ SFR tracer amounting to $\sigma_\alpha \equiv \sigma_{\rm SFR}/\sqrt{2} = 0.56, 0.24, 0.29\,$dex for elliptical, spiral and irregular galaxies, respectively. 

Adopting different boundaries between the three morphological categories does not improve the quality of the residuals. The latter show larger deviations from a standard normal distribution both i) when grouping transition-type S0/a galaxies ($T_{\rm LV}=0$) with (dwarf) elliptical and lenticular galaxies ($T_{\rm LV}<0$) rather than with spiral galaxies (KS-test p-value of 1\%), and ii) when grouping Sdm/Sc-Irr galaxies ($T_{\rm LV}=8$), dominated by a single spiral arm, with spiral galaxies rather than with irregular galaxies ($T_{\rm LV}>8$), with a marginal worsening of the KS test (p-value of 3\%). Including a further divide between early-type spirals and bulgeless disks at $T_{\rm LV}=5$ or between (dwarf) elliptical and lenticular galaxies at $T_{\rm LV}=-3$ does not improve the quality of the residuals. 

Figure~\ref{fig:intercal}, left, displays the best-fit $s_\alpha - s_{\rm FUV}$ relations. Elliptical galaxies show a nearly linear relation ($b\sim1$) and close to no offset ($a = 0.02\,$dex) between the two SFR tracers, as expected from galaxies with little ongoing star-formation activity. Irregular galaxies show a higher H$_\alpha$ to FUV ratio at higher SFR ($b>1$) suggesting enhanced star-formation activity on timescales of $5-10$\,Myr with respect to that on longer timescales. On the contrary, spiral galaxies show a lower H$_\alpha$ to FUV ratio at higher SFR ($b<1$) indicating that the most active episodes may have occurred more than 10\,Myrs ago for high-SFR spiral galaxies. 

Morphology may not be the only parameter driving star-forming activity. The morphological divide found in the present work may reflect the influence of, or be impacted by, other parameters such as the metallicity or the environment of the galaxies \citep{2018MNRAS.479.4136K}. The possible existence of hidden parameters is illustrated by the presence of outliers at the $4\,\sigma$ confidence level from the best-fit relation, which are responsible for the rather low KS-test p-value of 5\%.

A similar investigation is carried out by intercalibrating the H$_\alpha$ tracer with stellar mass. The following relations are found using $m_* = \log_{10} M_* / M_{\rm ref}$ with $M_{\rm ref} = 10^{9}\,M_\odot$

\begin{align}
\label{eq:m-sa}
    s_\alpha &= -3.28 + (1.09 \pm 0.12)\times m_*  & \text{ for } T_{\rm LV}<0, \nonumber\\
    &= -1.25 + (0.69 \pm 0.07)\times m_*  & \text{ for } 0\leq T_{\rm LV} \leq 7, \nonumber\\
    &= -1.15 + (1.21 \pm 0.04)\times m_* & \text{ for } T_{\rm LV}>7,
\end{align}
with a dispersion $\sigma_{\rm SFR} = 1.10, 0.41, 0.57\,$dex for elliptical, spiral and irregular galaxies, respectively.

The slopes $b_{m_*}(T_{\rm LV})$ obtained for the $s_\alpha - m_*$ relations are compatible with those obtained for the $s_\alpha - s_{\rm FUV}$ relations. The specific SFR, i.e.\ the SFR per stellar-mass unit, is nearly constant for elliptical galaxies, growing with increasing stellar mass for irregulars and decreasing with mass for spiral galaxies. As illustrated in Fig.~\ref{fig:intercal}, right, elliptical galaxies appear to be separated from main-sequence galaxies. The slopes of $1.21 \pm 0.04$ to $0.69 \pm 0.07$ inferred for irregular and spiral galaxies, respectively, with a crossing at about $10^{8.8}\,M_{\odot}$ are in line with slopes  in the literature ranging in $0.6-1.2$  along the main sequence \citep[see][and reference therein]{2019MNRAS.483.3213P}.
As for the $s_\alpha - s_{\rm FUV}$ relation, $4\,\sigma$ outliers result in a KS-test p-value for the residuals of 5\%, suggesting the possible presence of hidden parameters. 

The dispersion inferred for $s_\alpha$ in Fig.~\ref{fig:intercal}, left, is quadratically subtracted from $\sigma_{\rm SFR}$ inferred in Eq.~\ref{eq:m-sa}, so that the effective dispersion in the $M_*$-based estimator of SFR amounts to $\sigma_{m_*} = 0.94, 0.34, 0.49\,$dex, for elliptical, spiral and irregular galaxies, respectively. The uncertainties on the H$_\alpha$ SFR estimates from the three $s_\alpha - m_*$ relations are $40-70\%$ larger than those obtained from the corresponding $s_\alpha - s_{\rm FUV}$ relations, but the availability of morphological and stellar-mass information for a significant fraction of the revised MANGROVE sample largely makes up for the degraded SFR resolution.

\subsection{SFR of MANGROVE galaxies}\label{sec:sfr_function}

The LV sample lists 403 galaxies above the 2MPZ sensitivity limit with an estimated $s_\alpha$, labeled in the present work with \texttt{flag\_SFR=0}. For the other 277 LV galaxies, the $s_\alpha - m_*$ relations established in Eq.~\ref{eq:m-sa} as a function of morphology, $T_{\rm LV}$, provide an $s_\alpha$ estimate flagged with \texttt{flag\_SFR=1}. The inferred H$_\alpha$ SFR is larger (lower) than the upper (lower) limit determined by \cite{2018MNRAS.479.4136K} for 48 (2) LV galaxies, in which case the limit is used as putative $s_\alpha$, with \texttt{flag\_SFR=3} (\texttt{flag\_SFR=2}). Although providing SFR estimates for a majority of galaxies in the LV, this first approach leaves 99.9\% of galaxies in the revised MANGROVE sample without $s_\alpha$ estimates, when about one third of them have a morphological classification in the HyperLEDA database.

As discussed in Sec.~\ref{sec:LV}, the LV and HyperLEDA morphological classifications differ, particularly in their treatment of dwarf galaxies. Repeating the intercalibration analysis in Sec.~\ref{sec:sfr_calib} with $T_{\rm HL}$ instead of $T_{\rm LV}$ results in a worse separation of main-sequence and elliptical galaxies as well as worse residuals, with p-values for a Gaussian distribution that are at least an order of magnitude lower than those inferred with $T_{\rm LV}$, irrespective of the specific values adopted for the morphological boundaries. The morphological assignment according to HyperLEDA, $T_{\rm HL}$, rather than to the LV, $T_{\rm LV}$, is then disfavored to estimate $s_\alpha$. Instead, a mapping of $T_{\rm HL}$ as a function of $T_{\rm LV}$ for 583 LV galaxies above the 2MPZ sensitivity limit with both LV and HyperLEDA morphological classifications results in the following partitioning

\begin{equation}\label{eq:part_morph}
\hspace{-0.3cm}
\begin{footnotesize}
    \begin{bmatrix}
    \alpha_{\rm LV}(\rm E) \\ \alpha_{\rm LV}(\rm S) \\ \alpha_{\rm LV}(\rm Irr)
    \end{bmatrix}
    =
    \begin{bmatrix}
    0.771 && 0.073 && 0.156\\
    0.011 && f_{\rm S/S}(m_*) && f_{\rm S/Irr}(m_*)\\
    0.098 && 0.028 && 0.874
    \end{bmatrix}
    \begin{bmatrix}
    \alpha_{\rm HL}(\rm E) \\ \alpha_{\rm HL}(\rm S) \\ \alpha_{\rm HL}(\rm Irr)
    \end{bmatrix},
\end{footnotesize}
\end{equation}
where the matrix elements correspond to the average proportion of galaxies from each HyperLEDA morphological category, that is $T_{\rm HL}(\rm dE/E/S0)\leq -0.5$, $-0.5 < T_{\rm HL}(\rm S) \leq 7.5$, $7.5 < T_{\rm HL}(\rm Irr)$, associated to the three LV morphological categories.\footnote{The morphological classification on de Vaucouleurs' scale is provided as integers for $T_{\rm LV}$ while a finer distinction at the $0.1$ unit precision is provided for $T_{\rm HL}$.} The proportions are observed to be independent from stellar mass for HyperLEDA elliptical and irregular galaxies, which mostly display a confusion at the $10-15\%$ level between the two corresponding categories of the LV classification (see upper right and lower left corner of the matrix in Eq.~\ref{eq:part_morph}). Galaxies with $-0.5 < T_{\rm HL}(\rm S) \leq 7.5$ belong both to the spiral and irregular LV categories, with a larger proportion of irregulars at low masses and a higher proportion of spirals at high masses. This behavior is parameterized with $f_{\rm S/S}(m_*) = 0.989\times\mathcal{H}(\log_{10} M_*\,|\,8.5,0.5)$ and $f_{\rm S/Irr}(m_*) = 0.989\times\Big(1-\mathcal{H}(\log_{10} M_*\,|\,8.5,0.5)\Big)$, where $\mathcal{H}(x\,|\,\mu,\sigma^2) = \frac{1}{2}{\rm erfc}\Big(\frac{x-\mu}{\sqrt{2}\sigma}\Big)$ is the Gaussian cumulative distribution function.

The SFR of galaxies with an HyperLEDA morphological classification (\texttt{flag\_SFR=4}) is estimated as a weighted average of the SFRs of elliptical, spiral and irregular galaxies according to the $s_\alpha-m_*$ relations in Sec.~\ref{sec:sfr_calib}, with weights provided in the corresponding line of the matrix in Eq.~\ref{eq:part_morph}. For the remaining two thirds of galaxies in the revised MANGROVE sample with no morphological classification (\texttt{flag\_SFR=5}), the proportions of galaxies in the three HyperLEDA morphological categories, $\alpha_{\rm HL}$, are estimated from galaxies with \texttt{flag\_SFR=4}, that is a known HyperLEDA morphology, in eight logarithmic bins of distances between 0.02 and 350\,Mpc. The proportions obtained in the last bin between 100 and 350\,Mpc are used at larger distances, where the MANGROVE sample is not expected to be flux limited. The approach adopted to estimate the SFR of galaxies with $\texttt{flag\_SFR=5}$ assumes that they follow a similar  distribution as those with a morphological classification. Although individual SFR estimates are expected to be more uncertain for galaxies of unknown morphology, the first moment of their SFR distribution should by construction be representative of the average population, assuming no selection bias related to morphology.

The H$_\alpha$ SFR inferred from the $s_\alpha-m_*$ relations are based on LV estimates  by \cite{2018MNRAS.479.4136K} corrected for extinction, that is for escape of H$_\alpha$ emission from the galaxy, and on an H$_\alpha$ luminosity to SFR conversion factor of $8.79 \times 10^{-42}\,M_\odot\,{\rm yr}^{-1}/({\rm erg\,s}^{-1})$ \citep{1998ARA&A..36..189K}. As discussed in \cite{2003A&A...410...83H}, although largely independent from metallicity, such H$_\alpha$ estimates tend to underestimate the total SFR of galaxies because ionising photons processed into the H$_\alpha$ recombination line can be absorbed by dust grains beforehand. The total SFR based on H$_\alpha$ emission should then be estimated as 

\begin{equation}\label{eq:Ha_sfr_scaling}
 {\rm SFR} = C_{H_\alpha} \times L^{c}_{H_\alpha} /f,
\end{equation}
where $L^{c}_{H_\alpha}$ is the H$_\alpha$ luminosity corrected for extinction and $C_{H_\alpha} = 7.89 \times 10^{-42}\,M_\odot\,{\rm yr}^{-1}/({\rm erg\,s}^{-1})$ according to \cite{2003A&A...410...83H}. The latter factor is based on a Salpeter IMF and is converted to the reference Chabrier IMF, as indicated in introduction. The fraction of truly ionising radiation is inferred by \cite{2003A&A...410...83H} from joint UV, IR and H$_\alpha$ observations as $f = 0.57 \pm 0.21$, suggesting that ${\sim}\,40\% \pm 20\%$ of incoming ionising photons are absorbed instead of being processed into H$_\alpha$ emission. The SFR of all galaxies in the revised MANGROVE sample is thus based on Eq.~\ref{eq:Ha_sfr_scaling}, assuming $f=0.57$.

\begin{figure*}[ht!]
\epsscale{1.2}
\plotone{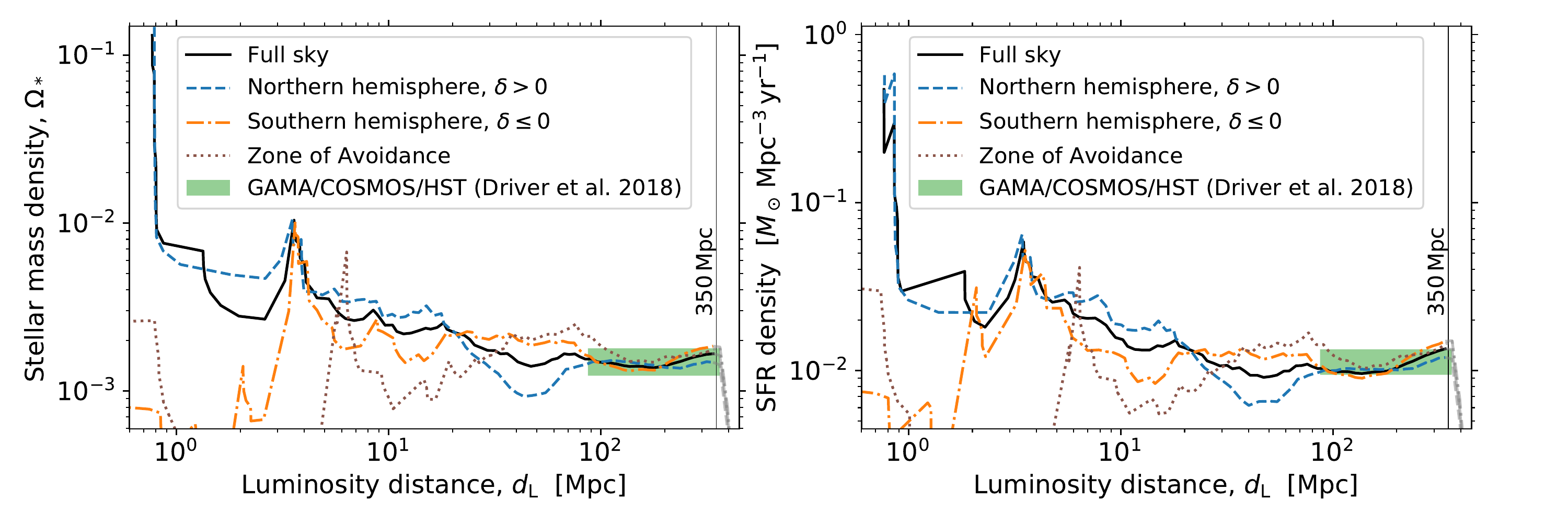}
\caption{The stellar-mass density (in critical units, \textit{left}) and star-formation-rate density (\textit{right}) in sliding distance windows of varying thickness, as detailed in the text. Density estimates are provided over the full-sky as black solid lines, in the Northern and Southern hemispheres as blue dashed and orange dash-dotted lines, respectively, and in the Zone of Avoidance as brown dotted lines. Density estimates are shown in gray beyond 350\,Mpc, where the sample is expected not to be flux limited. For reference, $68\%$ confidence-level intervals from deep-field observations in the local universe are shown as green bands.\label{fig:1D}}
\end{figure*}

The ionising fraction, $f=0.57$, shows a ${\sim}\,40\%$ uncertainty and depends on the assumed star-formation activity timescale \citep[$f = 0.39$ for 10\,Myrs instead of constant activity according to][]{2003A&A...410...83H} as well as on the sample of galaxies \citep[$f=0.77$ for a Salpeter IMF according to][]{2009ApJ...706.1527B}. The uncertainties on the SFRs of individual galaxies, estimated to be within a factor $2-3$ for irregulars and spirals with known morphologies according to Eq.~\ref{eq:m-sa},\footnote{The cross-match against the GAMA sample from \cite{2020MNRAS.498.5581B} suggests an asymmetric SFR resolution of $(-0.4,+0.8)\,$dex and a bias smaller than 0.05\,dex.} are expected to be washed out when averaged over a sufficiently large number of galaxies. Nonetheless, any SFR density estimated from the revised MANGROVE sample remains affected by an overall systematic uncertainty of ${\sim}\,40\%$ directly related to the absorption of ionising photons, which impacts all galaxies considered in the present work. 

As a first note, while H$_{\alpha}$ emission in dE/E/S0 galaxies may trace not only young stellar populations but also the presence of non-thermal emitters in the galaxy \citep{2015A&A...579A.102B}, elliptical galaxies are found to contribute to a negligible fraction of the SFR density in the local universe (see App.~\ref{app:SFR_dist}), which suggests a negligible bias induced by the $s_\alpha-m_*$ elliptical relation with respect to the overall systematic uncertainty on the ionizing fraction. As a second note, the relations established for spiral and irregular galaxies are expected to reproduce the average behavior of star-forming galaxies in and out of the main sequence, provided the Local Volume sample is representative of galaxies in the local universe. Nonetheless, SFR estimates may be under- or over-estimated at the individual galaxy level, e.g.\ for starburst galaxies outside of the main sequence \citep{2018A&A...616A.110E}. 

The SFR incompleteness as a function of Galactic latitude and distance is estimated following the methodology developed in Sec.~\ref{sec:mass}, as $c_{N, s}(d,b)= c_S(d)\times c_b(b)$ with $c_b(b)$ provided in Eq.~\ref{eq:galcorr} and where $c_S(d)$ accounts for the scaling relations in Eq.~\ref{eq:m-sa}:
\begin{equation}
\small
\label{eq:cS}
c_{S}(d) = \frac{\sum_{i, j = 1}^{(2, 3)}\phi_i\, \alpha_{\rm LV}(T_j)\, 10^{a(T_j)}\, \Gamma\big(1+\alpha_i+b(T_j), \frac{M_{\rm 2MPZ}(d)}{M_{\rm cut}}\big)}{\sum_{i, j = 1}^{(2, 3)}\phi_i\,  \alpha_{\rm LV}(T_j)\, 10^{a(T_j)}\, \Gamma\big(1+\alpha_i+b(T_j)\big)},
\end{equation}
where $j$ runs over the three morphological types, which are assumed to be distributed according to constant fractions, $\alpha_{\rm LV}$, in the local universe (see Appendix~\ref{app:SFR_dist}). The weights accounting for SFR completeness, $c_{N, s}(d,b)$, range in $[0.16,0.99]$ in $350\,{\rm Mpc}\times 4\pi$, so that the count weights of individual galaxies, $\{1/c_{S, m}(d_i,b_i)\}_i$,  do not exceed a factor of about six. 

\section{$M_*$ and SFR density out to 350\,Mpc} \label{sec:csm_csfr}

The revised MANGROVE sample includes the coordinates, distances, $M_*$, SFR estimates and completeness correction factors for 410,761 galaxies out to 350\,Mpc, 32,796 of which being clones in the ZoA.  Beyond 350\,Mpc, 99,311 galaxies, including 7,542 clones, are also listed, although the sample cannot be guaranteed to be flux limited beyond this limit (see Sec.~\ref{sec:mass}). The stellar-mass and SFR densities are visualized in 1D and 3D in the following sub-sections to investigate the consistency of the revised MANGROVE sample with cosmographies established in the local universe. A 2D visualization on the sphere further enables a comparison to astroparticle sky maps.

\subsection{1D visualization}\label{sec:1D}

The stellar-mass density (SMD) is estimated in sliding distance windows as $\frac{1}{V}\sum_i M_{*,\,i}/c_{N, m}(d_i,b_i)$, where $V$ is the comoving volume of interest and $d_i$, $b_i$ are the luminosity distance and Galactic latitude of the galaxy indexed by $i$. The mean distance of galaxies in each distance window is computed as a weighted average, with weights proportional to $M_{*}/c_{N, m}$. The same approach is applied to determine the SFR density (SFRD). The resulting SMD and SFRD obtained in windows of varying thickness are displayed in Fig.~\ref{fig:1D}. The thickness of the outer windows, whose geometric mean distance lies beyond 350\,Mpc, is taken as ${\pm}\,0.05\,$dex, matching the photometric distance resolution of 12\%. The innermost window is bounded by the Large Magellanic Cloud (LMC) and Andromeda galaxies, at 50 and 780\,kpc, respectively, which corresponds to a first-window thickness of ${\pm}\,0.6\,$dex. The thickness of intermediate distance windows is linearly decreasing with the logarithm of luminosity distance.

The SMD averaged over the full sky shows a behavior as a function of distance that is similar to that determined in \cite{2018AN....339..615K}, who combined the LV sample with the 2MRS-based catalog of galaxy groups from \cite{2015AJ....149..171T}. The SMD is dominated by the Local Group and in particular by Andromeda out to 1\,Mpc, by the Council of Giants at $3-6$\,Mpc \citep{2014MNRAS.440..405M} and by the Virgo Cluster at ${\sim}\,20\,$Mpc, before reaching a nearly constant value compatible with that inferred from the GAMA/G10-COSMOS/3D-HST survey at $0.02<z<0.08$ \citep{2018MNRAS.475.2891D}. The uncertainty on the latter SMD estimate is mostly attributed to cosmic variance. The SMD inferred from Northern- and Southern-hemisphere galaxies, at positive and negative declinations respectively, converge to the full-sky SMD beyond 100\,Mpc, suggesting a negligible North-South dipole farther away, that is an overall isotropy on large angular scales for the outermost layers. Notably, the SMD inferred in the ZoA also converges towards the full-sky value beyond $120\,$Mpc, confirming the effectiveness of the cloning procedure and of the correction for incompleteness at low Galactic latitudes (see Sec.~\ref{sec:ZoA}). The probed distance range extends to 350\,Mpc or 1 billion light-years, which can be compared to 135\,Mpc in \cite{2018AN....339..615K}. Beyond 350\,Mpc, the SMD sharply drops due to incompleteness, as indicated by the gray line portions in Fig.~\ref{fig:1D}.

The SFRD inferred out to 350\,Mpc displays a behavior similar to that of the SMD. The SFRD plateaus beyond 100\,Mpc at a value in good agreement with that inferred by \cite{2018MNRAS.475.2891D}. To the knowledge of the author, the results presented in Fig.~\ref{fig:1D}, right, constitute the first attempt to map the evolution of the SFRD out to a few hundreds of Mpc and over the entire sky.   

\begin{figure*}
\gridline{\hspace{-0.5cm}\fig{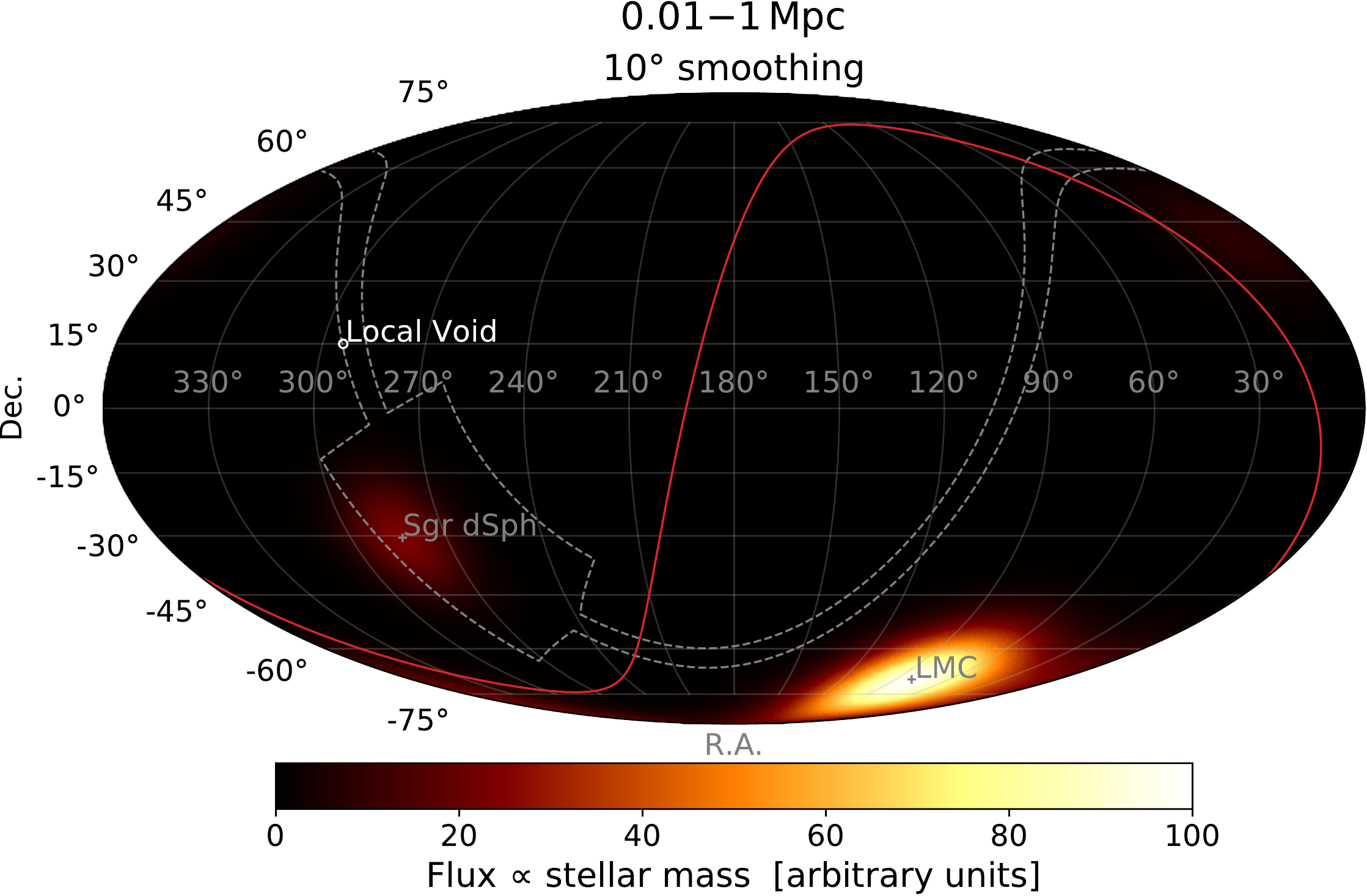}{0.49\textwidth}{\qquad(a)}
          \hspace{-0.5cm}\fig{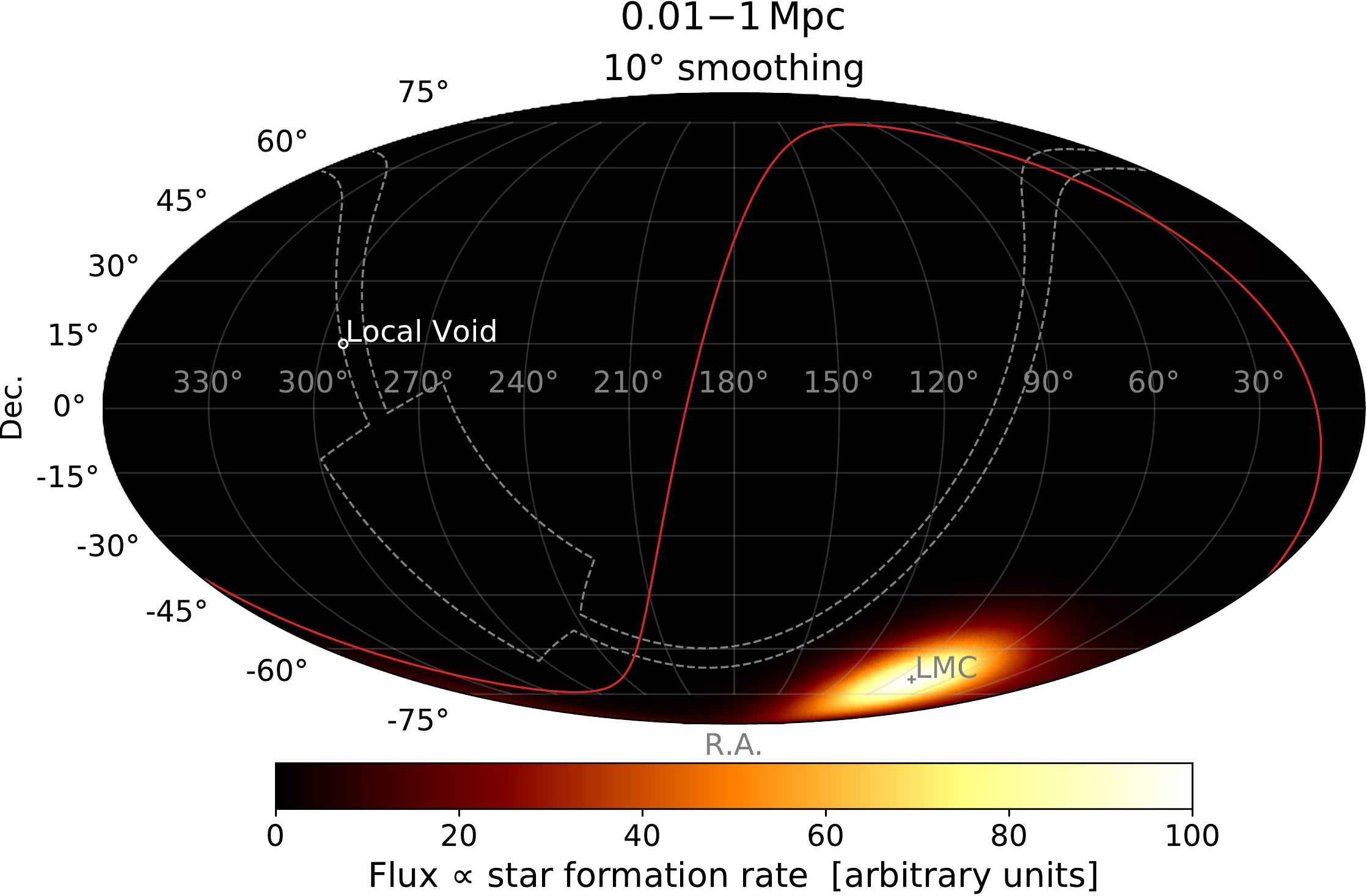}{0.49\textwidth}{\qquad(b)}}
\vspace{-0.25cm}
\gridline{\hspace{-0.5cm}\fig{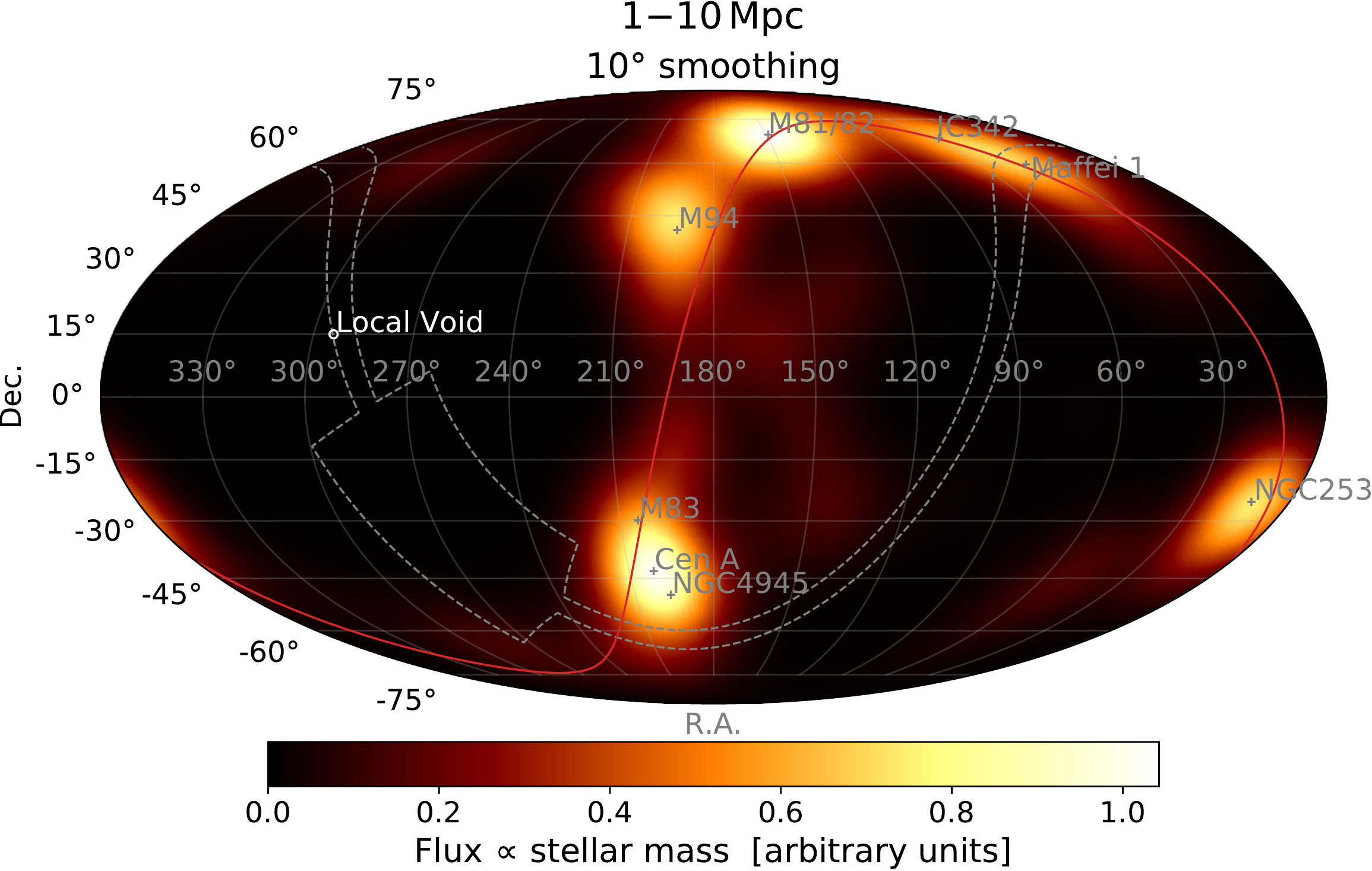}{0.49\textwidth}{\qquad(c)}
          \hspace{-0.5cm}\fig{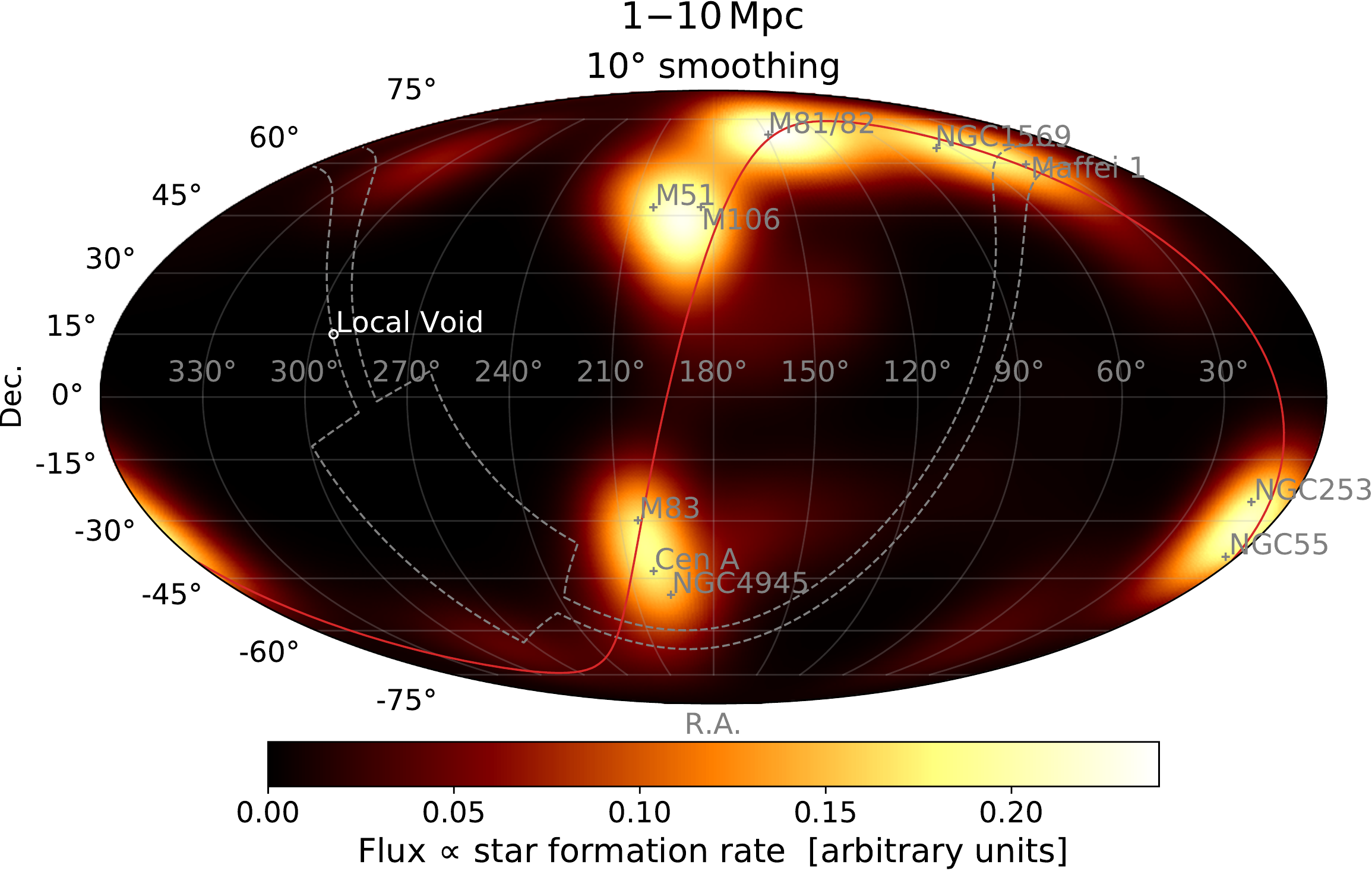}{0.49\textwidth}{\qquad(d)}}
\vspace{-0.25cm}
\gridline{\hspace{-0.5cm}\fig{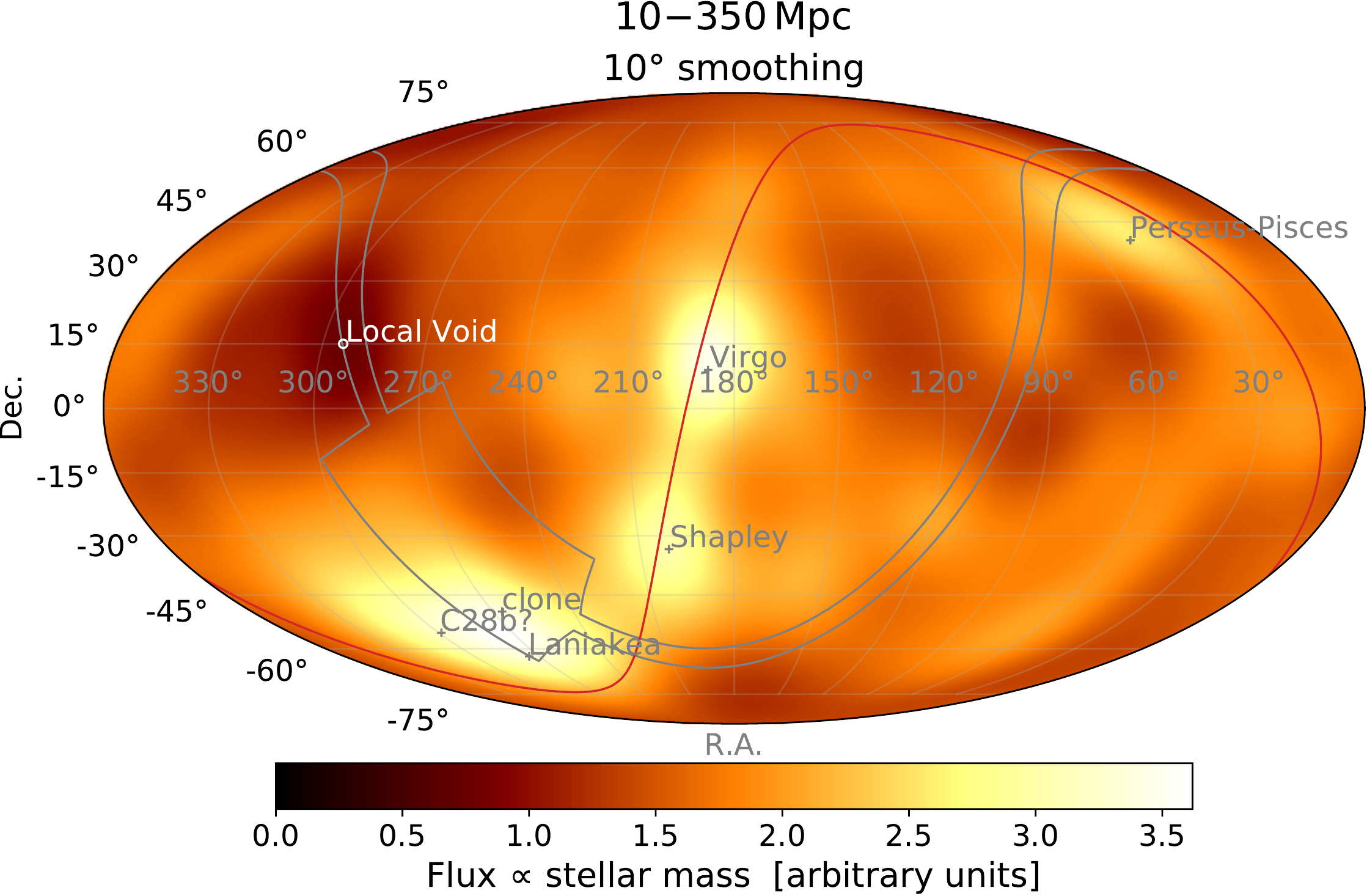}{0.49\textwidth}{\qquad(e)}
          \hspace{-0.5cm}\fig{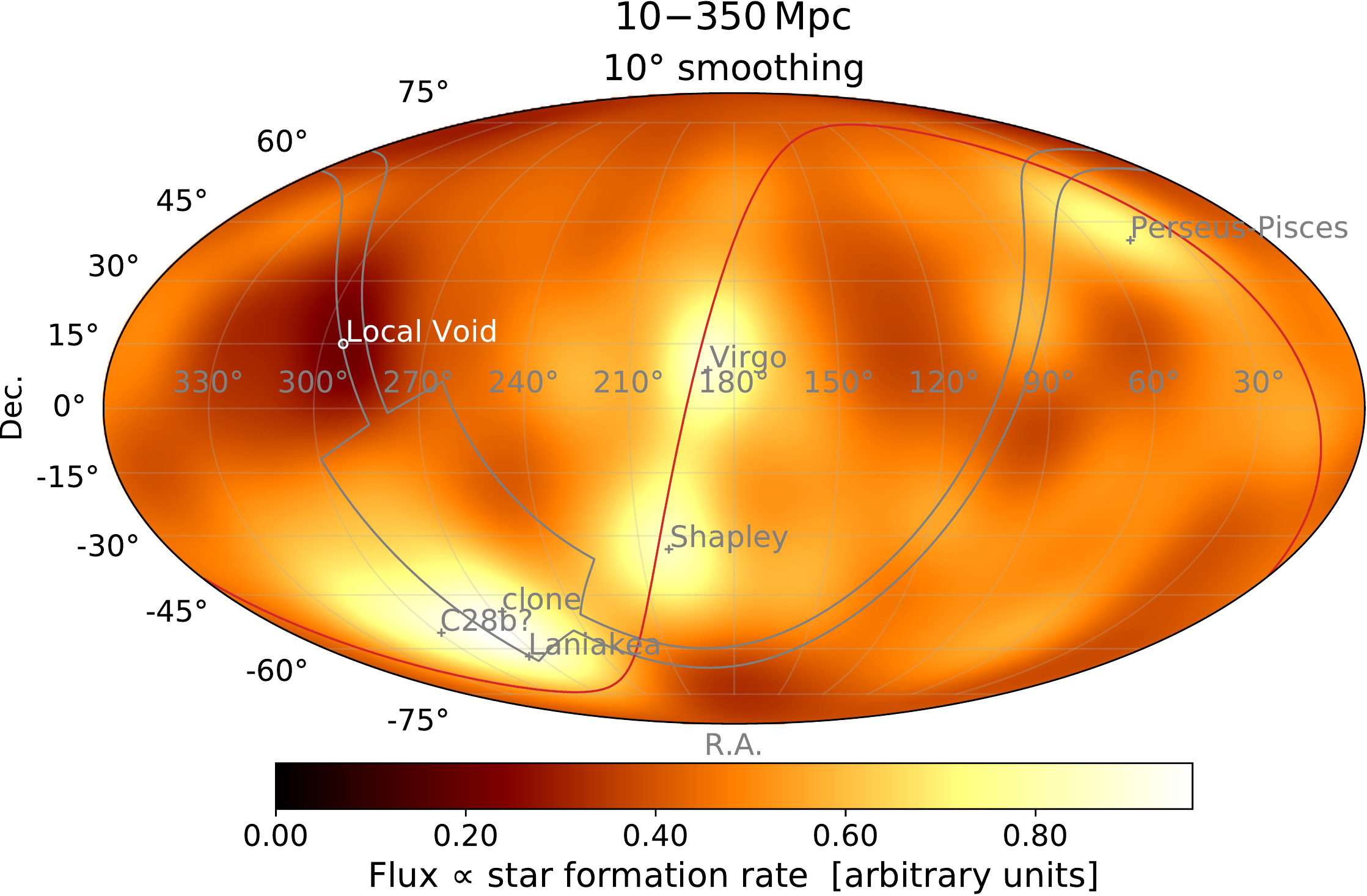}{0.49\textwidth}{\qquad(f)}}
\vspace{-0.25cm}
\caption{The sky maps, in equatorial coordinates, of the flux expected from galaxies at luminosity distances ranging in $0.01-1\,$Mpc (\textit{top}), $1-10\,$Mpc (\textit{mid}) and $10-350\,$Mpc (\textit{bottom}), in proportion to their stellar mass (\textit{left}) and star formation rate (\textit{right}). The flux maps are smoothed on a 10$^\circ$ angular scale. The color scale is normalized so that the peak fluxes in panels a and b reach one hundred arbitrary units. The largest flux excesses are flagged using the most prominent galaxy (\textit{top, mid}) or (super-)cluster (\textit{bottom}) in that direction, as labeled in gray text. The lowest flux point in the outer distance layer, associated to the Local Void, is marked as a white circle. The supergalactic plane is shown as a red solid line. The ZoA boundaries are marked with gray dashed (\textit{top, mid}) and solid (\textit{bottom}) lines, highlighting the application of ZoA correction and galaxy cloning at distances larger than 11\,Mpc. \label{fig:2D}}
\end{figure*}

\subsection{2D visualization}\label{sec:2D}

The flux from galaxies within 350\,Mpc is displayed on the sphere in Fig.~\ref{fig:2D}, for three distance ranges illustrating the relative contributions from the Local Group, the Council of Giants and local (super)clusters. The flux from galaxies is estimated in proportion to their stellar mass, $\frac{M_*/c_{N, m}(d,b)}{4\pi d^2}$, or SFR and normalized so that the peak flux of the innermost layer is set to one hundred arbitrary units. The flux maps are smoothed with a 10$^\circ$ von Mises–Fisher beam \citep{1953RSPSA.217..295F} to highlight flux overdensities on intermediate angular scales.

The first distance layer, covering the $0.01-1\,$Mpc distance range, is dominated by the nearby LMC ($d_{\rm L} \sim 50\,$kpc), with a smaller stellar-mass contribution in the ZoA from the Sagittarius Dwarf Spheroidal Galaxy (Sgr dSph, $d_{\rm L} \sim 20\,$kpc).\footnote{Smaller contributions from the Small Magellanic Cloud and Andromeda are barely visible in Fig.~\ref{fig:2D}, panel a. As a note, the contribution of Andromeda, which is 15 times more distance than the LMC, could be underestimated: its stellar mass is estimated at about $10^{10.5}\,M_\odot$ in the Local Volume sample, which can be compared to more advanced estimates in \cite{2012A&A...546A...4T}.} As a reminder, the galaxy cloning procedure presented in Sec.~\ref{sec:ZoA} is only applied beyond 11\,Mpc, which enables the visualization of contribution from local sources in the ZoA, at the expense though of a possible misestimate of the overall ZoA contribution in the Local Volume. The latter volume is nonetheless mostly composed of voids, as is apparent in panels c and d of  Fig.~\ref{fig:2D} that cover the $1-10\,$Mpc range. This distance range is dominated by the so-called Council of Giants  \citep{2014MNRAS.440..405M}, with galaxies distributed along the Local Sheet, that is a plane tilted by only $8^\circ$ from the supergalactic plane (see red line in Fig.~\ref{fig:2D}). Similar flux patterns are found using  SFR and stellar mass as flux proxies, with a Northern hemisphere brighter than the Southern one by $15\%$ and $30\%$, respectively. Voids bound the Local Sheet, with underdensities at right ascensions of about $120^\circ$ and $300^\circ$ along the celestial equator. The latter underdensity, known as the Local Void \citep{2008ApJ...676..184T}, shrinks but remains apparent in the distribution of galaxies at $10-350\,$Mpc, with a lowest flux at Equatorial coordinates $(\alpha, \delta) = (294^\circ, 15^\circ)$. As illustrated in panels e and f of  Fig.~\ref{fig:2D}, the structures dominating the flux from this outer layer are the Virgo cluster at a distance of ${\sim}\,20\,$Mpc \citep{2007ApJ...655..144M}, the Laniakea and Perseus-Pisces superclusters  \citep{2014Natur.513...71T}, with peak densities estimated in Sec.~\ref{sec:3D} to lie at $d_{\rm L} \sim 65-75\,$Mpc, and the more distant Shapley concentration discussed in  e.g.~\cite{2011MNRAS.416.2840L}, whose peak density is estimated to lie at $d_{\rm L} \sim 210\,$Mpc. The candidate supercluster identified as the C28 overdensity in \cite{2006MNRAS.373...45E} lies in the background of Laniakea at a distance estimated to $d_{\rm L} \sim 210-260\,$Mpc. As for Laniakea that is close to the ZoA, the cloning procedure could affect the relative flux estimated from C28. 

\subsection{3D visualization}\label{sec:3D}

Finally, a 3D visualization of SMD within 350\,Mpc (${\sim}\,350\,$Mpc in comoving units) is presented in Fig.~\ref{fig:3D}, with zoomed-in interactive versions available in Appendix~\ref{app:3d}. As in Fig.~\ref{fig:1D} and \ref{fig:2D}, galaxy clones in the ZoA and corrections for incompleteness are accounted for. The Milky Way is included in the volume following the stellar-mass and SFR estimates from \cite{2015ApJ...806...96L}, converted from a Kroupa to a Chabrier IMF. The density field is smoothed with a 3D Gaussian filter and iso-density contours are shown as colored surfaces at $20\%-80\%$ of the maximum density in the volume, by steps of 10\%.

\begin{figure*}
\epsscale{1.2}
\plotone{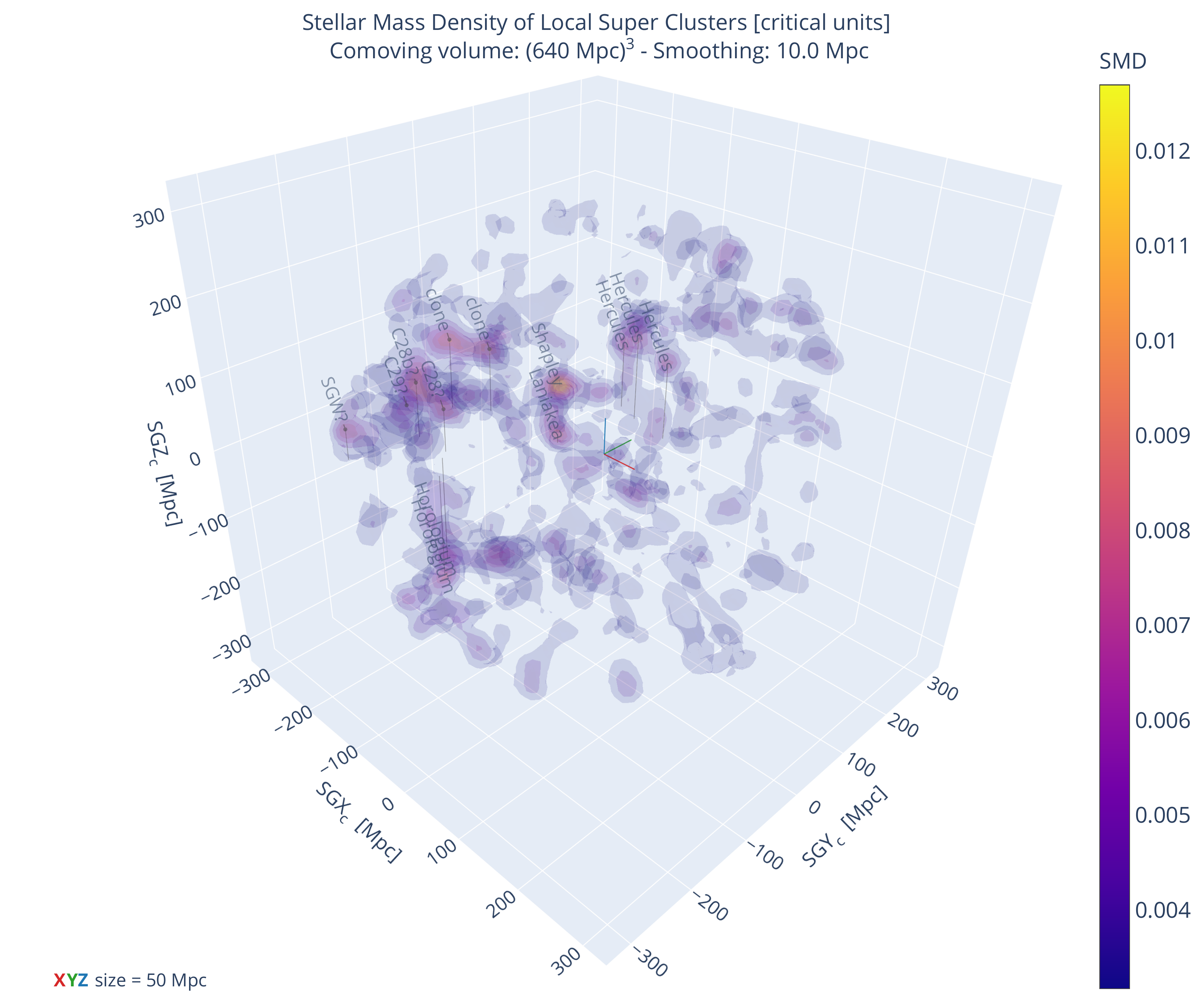}
\caption{The stellar-mass density in a 640\,Mpc-side cube, in comoving supergalactic coordinates. The density field is smoothed with a 3D Gaussian filter on a comoving scale of 10\,Mpc. Seven iso-density contours ranging in $20-80\%$ of the maximum are illustrated with varying colors. Prominent objects with peak densities larger than half of the maximum density in the volume are listed. Thin vertical gray lines indicate the distance to the supergalactic plane. Acronyms denote the Southern Great Wall (\texttt{SGW?}) and the C29/C28 (\texttt{C29?}, \texttt{C28?}, \texttt{C28b?}) overdensities from \cite{2006MNRAS.373...45E}, including clones of C28 in the ZoA. 
\label{fig:3D}}
\end{figure*}

Zoomed-in visualizations of the Local Volume in Appendix~\ref{app:3d} encompass a sphere of $6.25\,$Mpc-radius, which is identified in \cite{2014MNRAS.440..405M} as the boundary of the Council of Giant. The planar distribution of this ``council'' of galaxies along the Local Sheet is more easily seen in the interactive version of the figure. Galaxies with a high H$_\alpha$ SFR, namely M\,101 (so-called Pinwheel galaxy), NGC\,6946 (so-called Fireworks galaxy) and NGC\,3621 at luminosity distances of $6-7\,$Mpc are also found $2-4\,$Mpc above and below the Local Sheet. 

The Virgo cluster dominates the distribution of SMD and SFRD in a comoving volume of 50\,Mpc side. The elongated shape of the Virgo cluster away from the Milky Way (see interactive version) and its peak density location, at cartesian supergalactic coordinates $(-4.5, +15.5, -1.0)\,$Mpc on a 0.5\,Mpc grid, agree well with those reconstructed by \cite{2007ApJ...655..144M} in a dedicated HST survey of early-type galaxies. The Fornax cluster \citep{2007ApJS..169..213J} is, as expected, identified as the second densest structure at $(-2.0, -15.0, -13.5)\,$Mpc. A peak SMD larger than half that of Virgo is also found for a third group, whose most massive galaxy is identified as PGC\,054411. This barred spiral galaxy, with an estimated stellar mass $M_* = 10^{11.16}\, M_\odot$, lies close to the ZoA at $b \sim 4^\circ$. The correction factor for incompleteness enhances its contribution to the local density by 60\%. The contribution of PGC\,054411 is further doubled since its clone is added in the ZoA. PGC\,054411 and its neighbors are not included in the catalog of galaxy groups by \cite{2015AJ....149..171T} but PGC\,054411 is present in the V8k catalog investigated by \cite{2013AJ....146...69C}. Further studies of the velocity dispersion of galaxies in this region would be desirable to determine whether the overdensity around PGC\,054411 is real or a mere artifact of the galaxy cloning approach. 

The density fields smoothed on a 10\,Mpc scale out to comoving distances of 300\,Mpc are also available in Appendix~\ref{app:3d}. The densest structure in the volume is the Shapley Concentration, whose peak on a 5\,Mpc grid is located at $(-175, +100, -10)\,$Mpc, corresponding to a luminosity distance of 210\,Mpc. This peak is located less than 9\,Mpc away from the main attractor identified in the 3D velocity field of Cosmicflows-2 \citep{2017NatAs...1E..36H}. Our own supercluster, Laniakea, shows a peak density located at $(-60, -20, +10)\,$Mpc on a 5\,Mpc grid, that is shifted by ${\sim}\,10\,$Mpc out to $(-55, -30, +10)\,$Mpc when discarding clones in the ZoA. Both locations correspond to a luminosity distance of 65\,Mpc. These peak positions appear to be consistent with the overdensity fields displayed in \cite{2019MNRAS.489L...1D} but lie ${\sim}\,3\,\sigma$ away from the corresponding velocity-field attractor at $(-58\pm6, +6\pm10, +21\pm15)\,$Mpc/$h$. A smaller grid of 2.5\,Mpc with a 5\,Mpc-scale smoothing suggests the presence of two sub-peaks in the Laniakea supercluster, which are located at $(-55.0, -25.0, +12.5)\,$Mpc or $d_{\rm L} = 65\,$Mpc for the densest one and at $(-72.5, -10.0, +7.5)\,$Mpc or $d_{\rm L} = 75\,$Mpc for the lightest of the two. The X and Y supergalactic coordinates of the densest peak, closer to the ZoA, are affected by ${\pm}\,2.5\,$Mpc offsets if clones are discarded. The companion of Laniakea, that is the Perseus-Pisces supercluster, is found to peak at $(+65, -25, -15)\,$Mpc with a 10\,Mpc smoothing and at $(+70.0, -17.5, -17.5)$ with a 5\,Mpc smoothing. These peak locations appear, as for Laniakea, to be consistent with the overdensity field presented in \cite{2019MNRAS.489L...1D} but are ${\sim}\,5\,\sigma$ away from the corresponding velocity-field attractor at $(+33\pm10, -7\pm10, +31\pm9)\,$Mpc/$h$. 

Two major overdensities are identified in the background of Laniakea, showing directional coincidence with the C28 and C29 overdensities identified with 2MASS data by \cite{2006MNRAS.373...45E}. The structure possibly associated to C28, which is close to the ZoA, displays two peaks denoted as \texttt{C28?}\ and \texttt{C28b?}\ in Fig.~\ref{fig:3D}, which are located at $(-140, -135, +55)\,$Mpc and $(-190, -135, +75)\,$Mpc and correspond to luminosity distances of 210 and 260\,Mpc, respectively. The clones of these two peaks appear as mirrored structures in the ZoA and are responsible for the enhanced flux along the inner Galactic Plane in the lower panels of Fig.~\ref{fig:2D}. The more distance structure possibly associated to the C29 overdensity from \cite{2006MNRAS.373...45E}, denoted as \texttt{C29?}\ in Fig.~\ref{fig:3D}, shows a peak at $(-275, -85, -10)\,$Mpc, corresponding to a luminosity distance of 310\,Mpc.\footnote{The bottom panels of Fig.~\ref{fig:3D} also include a structure labeled as the Southern Great Wall (\texttt{SGW?}) at $(-205, -230, 40)\,$Mpc. This structure named by \cite{2006MNRAS.373...45E} differs from the Southern Wall at $(30, -70, -10)\,$Mpc neighbouring Perseus-Pisces and from the Great Wall, which connects Coma at $(0, 100, 10)\,$Mpc to one the three Hercules overdensities at $(-65, 105, 90)\,$Mpc \citep[see interactive version and][]{2013AJ....146...69C}.} As for the overdensity identified around PGC\,054411, studies of the velocity field in the region of C28 and C29 would help confirming the reality of these structures. Fruitful results could be expected from a comparison of structures identified in the present work with those traced with cosmic flows, particularly near the ZoA \citep[e.g.][]{2017MNRAS.466L..29K, 2017ApJ...845...55P, 2020ApJ...897..133P}.

\section{Discussion and conclusion}

The gravitational-wave community built the GLADE catalog as an assembly of two main components: the 2MPZ catalog, containing over 900,000 photometric distance estimates, and the 2MRS catalog, containing a subset of about 45,000 spectroscopic redshifts. The community further supplemented the GLADE assembly with stellar-mass estimates from WISE to construct the MANGROVE sample of about 800,000 galaxies. The present work proceeded in determining the flux limit of the GLADE and MANGROVE assemblies and in demonstrating that the stellar-mass distribution of galaxies above the 2MPZ sensitivity is consistent with the mass function estimated in deep-field observations, except for a high-mass tail that could deserve further scrutiny. With a flux-limited sample at hand, incompleteness as a function of distance and of Galactic latitude is corrected for over 90\% of the sky, while the remaining 10\% are filled through galaxy cloning along the Galactic plane. 

Over 150,000 distances have been revised with spectroscopic estimates from the HyperLEDA database and a particular attention was paid to galaxies in the Local Group and Local Sheet, by including information from the Local Volume catalog. The revised MANGROVE sample includes over 400,000 galaxies out to 350\,Mpc that are used to build up a cosmography of stellar mass. The scaling relation of H$_\alpha$ emission with stellar mass of galaxies with a known morphology in the Local Volume further enables a statistical estimate of the SFR of galaxies out to 350\,Mpc, based on either their observed morphology (about one third of the sample) or inferred from the overall morphological distribution of galaxies at comparable distances (remaining two thirds). Both SFRD and SMD show a plateau at distances larger than 100\,Mpc, at values consistent with those inferred from deep-field observations. The convergence of the SFRD and SMD in the outer distance layers, as estimated from the Northern and Southern hemispheres, suggests that the densities could be extrapolated at larger redshifts using an isotropic component following the cosmic evolution of the two tracers \citep[e.g.][]{2014ARA&A..52..415M}. 

Flux maps based on stellar-mass and SFR tracers illustrate the dominance over the extragalactic sky of very-nearby galaxies in the Local Group, in particular the LMC. The peak flux expected from galaxies in the Local Sheet corresponds to $0.2-1\%$ of that expected from galaxies in the Local Group, depending on the adopted tracer. The Local Sheet emission shows a flux pattern similar to the one that is emerging beyond the toe of the UHECR spectrum \citep{2019EPJWC.21001005B}. This possible agreement has already been quantified by the Pierre Auger and Telescope Array collaborations through analyses of arrival directions above 40\,EeV against a catalog of a few dozen nearby galaxies with a high SFR, as traced by their radio emission \citep{2018ApJ...853L..29A, 2018ApJ...867L..27A}. The source model adopted in these works, which is in good agreement with the flux map derived from over 800 galaxies between 1 and 10\,Mpc in the revised MANGROVE sample, led to evidence for anisotropy at a confidence level of $4\,\sigma$ in the largest UHECR datasets studied to date \citep{2018ApJ...853L..29A,2019arXiv190909073T}. Above 40\,EeV, the suppression of the ${\sim}\,3$ times brighter background from galaxies between 10 and 350\,Mpc could be understood from UHECR propagation in the cosmic microwave and infrared backgrounds \citep{2013APh....41...94A}. Nonetheless, if UHECR emission is traced by SFR or stellar mass, the contribution from the foreground layer that is the Local Group would need to be reduced by two-to-three orders of magnitude not to dominate the UHECR sky. To tackle this point, a follow-up study will investigate the impact of propagation in magnetic fields on Mpc scales in the context of a transient origin of UHECRs. For low magnetic-field column densities, as expected for very-nearby galaxies, the effective UHECR transient duration observed at Earth should be similar to the transient duration at escape from the host galaxy, while the temporal spread induced by propagation in a local magnetic field on Mpc scales would increase quadratically with the distance of host galaxies farther away \citep[e.g.][]{2008PhRvD..78b3005M}.\footnote{The temporal spread increases quadratically with distance for small angular deviations induced by a coherent magnetic field embedding the source and the observer. The evolution with distance is cubic in the regime of large angular deviations. See App.~E in \cite{2008PhRvD..78b3005M}.} Transients from very-nearby galaxies would thus be missed most of the time because of their shorter duration while more distant bursts could appear as persistent afterglows in the UHECR sky, provided a sufficiently high rate of UHECR bursts per SFR or stellar-mass unit. 

With a reach out to one billion light-years, the revised MANGROVE sample should encompass host galaxies of UHECR transients down to the instep region of the UHECR spectrum, located by the Pierre Auger Collaboration at $10-15\,$EeV. Provided a close-to-isotropic distribution of more distant galaxies, a successful transient UHECR model based on SFR should then be able to reproduce the dipole discovered by the Pierre Auger Collaboration above 8\,EeV. The potential of such a model to reproduce the large-scale UHECR anisotropy is supported by the directional coincidence of the Local Void with the UHECR underdensity at Equatorial coordinates $(\alpha, \delta) \approx (300^\circ, 15^\circ)$ \citep{2017Sci...357.1266P, 2018ApJ...868....4A}. Investigations of density fields expected from cosmic flows against the UHECR sky have already shown promising results on large angular scales \citep{2019MNRAS.484.4167G, 2021arXiv210104564D}. The present work further offers the possibility to jointly model the UHECR sky above the ankle and that beyond the toe of the UHECR spectrum, by resolving structures ranging from galaxies in the Local Sheet out to distant superclusters. 

As illustrated in the flux maps from galaxies at $10-350\,$Mpc, clusters such as Virgo would be expected to provide a non-negligible contribution to the UHECR sky above the ankle. Whether this expectation can be reconciled with a UHECR dipole pointing $90^\circ$ away from Virgo remains to be determined. A point that may not have received sufficient attention from the UHECR community lies in the confinement of UHECRs within intracluster magnetic fields \citep[e.g.][]{2008PhRvD..78b3005M,2018NatPh..14..396F}. At $E/Z = 5\,$EV, which is the cutoff rigidity inferred from the joint modeling of UHECR spectral and composition data, the UHECR gyroradius in a $5\,\mu$G field is 

\begin{equation}
\label{eq:Larmor}
r_{\rm L} \sim 1{\rm\,kpc} \times \Big(\frac{E/Z}{5{\rm\,EV}}\Big)\Big(\frac{B}{5\,\mu{\rm G}}\Big)^{-1}.     
\end{equation}

A reference magnetic-field strength of $5\,\mu$G is used in Eq.~\ref{eq:Larmor} as it corresponds to the peak value of the Coma intracluster magnetic field, whose coherence length is $\lambda \sim 10\,$kpc \citep{2010A&A...513A..30B}. With gyroradii, $r_{\rm L},$ at least an order of magnitude smaller than the intracluster coherence length, $\lambda$, virtually no UHECR should escape from the most magnetized clusters. The magnetic fields in clusters and filaments are expected to result from the amplification of a seed field through adiabatic compression of baryonic matter \citep{2018SSRv..214..122D}, providing a direct link to the 3D mapping of stellar mass established in the present work. This link will be exploited in the follow-up study of a transient UHECR scenario to infer intracluster magnetic fields and to determine which structures are switched off by magnetic confinement. 

The revised MANGROVE sample could serve other purpose than the cartography of the UHECR sky. On the one hand, the corrections developed in the present work enable an unbiased estimate of stellar-mass and SFR densities as a function of distance. Such corrections could be of use for gravitational-wave events with large error boxes, in particular those with the most uncertain distance estimates \citep[see][for an updated overview of gravitational-wave events]{2019PhRvX...9c1040A,2020arXiv201014527A}, and to establish priors on the distribution of matter that enable constraints on $H_0$ from observations of binary black-hole mergers \citep{2021ApJ...909..218A}. On the other hand, the supplementing of over 150,000 spectroscopic redshifts above the 2MPZ sensitivity could ease the identification of the host galaxies and their follow-up observations. Other applications to astroparticle physics and cosmology, discussed in Sec.~\ref{sec:intro}, include cross correlations with TeV-PeV neutrinos, constraining the sources of the extragalactic gamma-ray background, including possible dark-matter decay \citep{2017ApJS..232...10C,2018PhRvD..98j3007A}, and determining the stellar mass and SFR of galaxy groups, clusters and superclusters within a billion light-year.

A natural extension of the present work would consist in complementing the revised MANGROVE sample, or similarly a 2MPZ-based catalog of stellar mass \citep[see e.g.][for alternative stellar-mass estimators]{2014ApJ...782...90C}, with photometric and spectroscopic distances down to the WISE$\times$SCOS sensitivity limit. Using the 2MPZ catalog, the present work provides access to galaxies about seven times dimmer and $2.5-3$ times more distant than those listed in 2MRS. A census of galaxies a hundred times dimmer and ten times more distant than those in 2MRS could be reached by exploiting the full potential of WISE through WISE$\times$SCOS photometric redshifts. Nonetheless, particular attention should be paid to the regions excluded in WISE$\times$SCOS, around the Zone of Avoidance and in extragalactic areas dominated by bright foreground sources. Efforts to improve the completeness of the near-IR sky close to the Galactic Plane \citep[e.g.][]{2019ApJS..245....6M} will help in reducing the area where galaxy cloning is employed. Further extensions of the present work could exploit the fitting of spectral energy distributions of galaxies \citep[e.g.][]{2016ApJS..227....2S}, as available in optical to near-IR bands from 2MPZ entries. Provided sufficient multi-wavelength coverage, morphological assignment based on color-color selection and deviations from a linear light-to-mass scaling relation could also be investigated \citep[see e.g.][]{2010ApJ...709..644I}.  Supplementing near-IR observations with UV data from GALEX and far-IR data from IRAS would probably be needed to obtain, through fitting of spectral energy distributions, reliable estimates of stellar mass, SFR and possibly star-formation history of individual galaxies, as well as quantitative estimates of the associated uncertainties. The ongoing eROSITA survey \citep{2021A&A...647A...1P} could finally benefit such endeavors, particularly by providing a tracer for non-thermal activity that might contaminate or even outshine stellar emission from the host galaxy. 

Encouraging prospective for astrophysics, cosmology and astroparticle physics is expected as the various communities interested in stellar emission from extragalactic objects join forces and expertise.

\acknowledgments

The author is indebted to Dmitry Makarov for providing help with HyperLEDA queries and for discussions on the Local Volume sample, to Maciej Bilicki for exchanges on the 2MPZ and WISE$\times$SCOS catalogs, to Emeric Le Flo'ch and Nicolas Leroy for feedback on the MANGROVE sample, to Sabine Bellstedt, Simon Driver and the GAMA team for sharing stellar-mass and SFR estimates in the GAMA field, to Remi Adam for exchanges on cluster magnetic fields, to Toshihiro Fujii for discussions on voids and the UHECR sky, as well as to Olivier Deligny, Peter Tiniakov and Armando di Matteo for insightful exchanges on transient UHECR scenarios. The author gratefully acknowledges comments on this manuscript from Armando di Matteo, Remi Adam, Daniel Pomar\`ede, Emeric Le Fl'och, Olivier Deligny and Maciej Bilicki. The author is grateful to the anonymous referee for suggestions which improved the consistency of the manuscript.

This research has made use of the SIMBAD database, operated at CDS, Strasbourg, France. This research has made use of the NASA/IPAC Extragalactic Database (NED),
which is operated by the Jet Propulsion Laboratory, California Institute of Technology, under contract with the National Aeronautics and Space Administration. We acknowledge the usage of the HyperLeda database (http://leda.univ-lyon1.fr).

\software{astropy \citep{2013A&A...558A..33A}, plotly (https://plot.ly)}

\newpage
\appendix

\section{Revised MANGROVE sample}\label{app:revised_sample}

The first 25 entries of the revised MANGROVE sample are shown in Table~\ref{tab:rev_man}. The first column indicates the name of each galaxy, provided as ``clone'' in case of galaxies mirrored in the Zone of Avoidance. The next three columns provide the identifier from the AllWISE catalog, the IDX from GLADE and the PGC number from HyperLEDA. The next two columns provide the Equatorial coordinates of each galaxy, followed by an association flag: 1 and 2 for robust and tentative angular associations, respectively. The next two columns indicate the revised luminosity distance in Mpc and the associated flag, either 1 for 2MPZ photometric estimates, 2 or 3 for spectroscopic estimates only included in GLADE, 4 for LV distances, 5 and 6 for spectroscopic and cosmic-ladder estimates from HyperLEDA, respectively. The next three columns provide the revised stellar mass, in solar mass units, the associated flag (0 and 1 for estimates from the LV and MANGROVE samples, respectively) and the stellar-mass incompleteness correction factor accounting for proximity to the Galactic plane and for the 2MPZ sensitivity. The next three columns provide a statistical estimate of the total SFR of each galaxy, normalized to 1\,$M_\odot/$yr, the associated flag (0 for an H$_\alpha$-based estimate, 1 for objects with LV morphology, 2 and 3 for objects with H$_\alpha$ lower and upper limits, respectively, 4 for objects with  HyperLEDA morphology, 5 for objects with no morphological type) and the SFR incompleteness correction factor. Table~\ref{tab:rev_man}, as well as Tables~\ref{tab:mangrove_pgc}-\ref{tab:excluded} and a loading script enabling a 3D visualization of the Laniakea and Perseus-Pisces superclusters, are available at \url{http://doi.org/10.5281/zenodo.5118697}.

\begin{changemargin}{24cm}{1cm}

\begin{longrotatetable}
\begin{footnotesize}

\begin{deluxetable}{ccrrrrcrcrcrrcr}
\tablecaption{Revised MANGROVE sample above the 2MPZ sensitivity limit including clones in the Zone of Avoidance\label{tab:rev_man}}
\tablehead{\colhead{Name} & \colhead{AllWISE id} & \colhead{idx} & \colhead{pgc} & \colhead{R.A.} & \colhead{Dec} & \colhead{flag\_asso} & \colhead{$d_{\rm L}$} & \colhead{flag\_d} & \colhead{$\log_{10} M_*$} & \colhead{flag\_M*} & \colhead{$c_{\rm N, m}$} & \colhead{$\log_{10} {\rm SFR}$} & \colhead{flag\_SFR} & \colhead{$c_{\rm N, s}$}\\ \colhead{$\mathrm{}$} & \colhead{$\mathrm{}$} & \colhead{$\mathrm{}$} & \colhead{$\mathrm{}$} & \colhead{$\mathrm{{}^{\circ}}$} & \colhead{$\mathrm{{}^{\circ}}$} & \colhead{$\mathrm{}$} & \colhead{$\mathrm{Mpc}$} & \colhead{$\mathrm{}$} & \colhead{[$M_\odot$]} & \colhead{$\mathrm{}$} & \colhead{$\mathrm{}$} & \colhead{[$M_\odot$/yr]} & \colhead{$\mathrm{}$} & \colhead{$\mathrm{}$}}
\startdata
NGC0253 & J004733.14-251717.7 & 0 & 2789 & 11.893 & -25.292 & 1 & 3.7 & 4 & 10.76 & 0 & 0.997 & 0.26 & 0 & 0.913 \\
NGC5128 & J132527.62-430109.1 & 1 & 46957 & 201.370 & -43.017 & 1 & 3.68 & 4 & 10.67 & 0 & 0.997 & -0.06 & 0 & 0.913 \\
NGC5236 & J133700.61-295155.5 & 3 & 48082 & 204.250 & -29.868 & 1 & 4.90 & 4 & 10.64 & 0 & 0.996 & 0.44 & 0 & 0.903 \\
NGC4736 & J125053.14+410712.7 & 4 & 43495 & 192.723 & 41.119 & 1 & 4.41 & 4 & 10.34 & 0 & 0.996 & -0.43 & 0 & 0.907 \\
NGC0055 & J001453.92-391156.4 & 5 & 1014 & 3.785 & -39.220 & 1 & 2.11 & 4 & 9.26 & 0 & 0.998 & -0.36 & 0 & 0.930 \\
NGC0300 & J005453.46-374103.0 & 6 & 3238 & 13.723 & -37.682 & 1 & 2.09 & 4 & 9.19 & 0 & 0.998 & -0.78 & 0 & 0.930 \\
NGC5102 & J132157.60-363748.3 & 7 & 46674 & 200.491 & -36.630 & 1 & 3.66 & 4 & 9.48 & 0 & 0.997 & -1.99 & 0 & 0.913 \\
NGC7793 & J235749.73-323527.5 & 8 & 73049 & 359.456 & -32.590 & 1 & 3.63 & 4 & 9.48 & 0 & 0.997 & -0.45 & 0 & 0.914 \\
NGC4548 & J123526.45+142946.9 & 9 & 41934 & 188.860 & 14.496 & 1 & 15.62 & 6 & 10.54 & 1 & 0.987 & 0.00 & 4 & 0.850 \\
NGC6503 & J174926.45+700840.8 & 10 & 60921 & 267.365 & 70.145 & 1 & 6.25 & 4 & 9.78 & 0 & 0.995 & -0.37 & 0 & 0.894 \\
NGC4442 & J122803.90+094813.3 & 11 & 40950 & 187.016 & 9.804 & 1 & 15.12 & 6 & 10.51 & 1 & 0.988 & 0.11 & 4 & 0.852 \\
NGC4469 & J122928.05+084500.8 & 13 & 41164 & 187.367 & 8.750 & 1 & 16.75 & 6 & 10.30 & 1 & 0.986 & -0.16 & 4 & 0.846 \\
NGC0247 & J004707.60-204529.9 & 14 & 2758 & 11.785 & -20.76 & 1 & 3.72 & 4 & 9.28 & 0 & 0.997 & -0.55 & 0 & 0.913 \\
NGC4586 & J123828.39+041909.0 & 15 & 42241 & 189.618 & 4.319 & 1 & 13.06 & 6 & 9.84 & 1 & 0.989 & -0.47 & 4 & 0.860 \\
NGC4440 & J122753.56+121735.8 & 17 & 40927 & 186.973 & 12.293 & 1 & 16.73 & 6 & 9.96 & 1 & 0.986 & -0.39 & 4 & 0.846 \\
NGC4342 & J122338.99+070314.4 & 18 & 40252 & 185.912 & 7.054 & 1 & 12.61 & 5 & 9.66 & 1 & 0.990 & -0.88 & 4 & 0.862 \\
NGC4808 & J125548.92+041814.8 & 19 & 44086 & 193.954 & 4.304 & 1 & 17.43 & 6 & 9.99 & 1 & 0.986 & -0.37 & 4 & 0.844 \\
NGC4387 & J122541.67+124838.1 & 20 & 40562 & 186.424 & 12.811 & 1 & 17.76 & 6 & 9.89 & 1 & 0.985 & -0.62 & 4 & 0.843 \\
NGC4424 & J122711.57+092513.9 & 21 & 40809 & 186.798 & 9.421 & 1 & 4.14 & 6 & 8.65 & 1 & 0.997 & -1.32 & 4 & 0.909 \\
NGC4458 & J122857.55+131430.8 & 22 & 41095 & 187.240 & 13.242 & 1 & 16.31 & 6 & 9.72 & 1 & 0.987 & -0.81 & 4 & 0.848 \\
NGC4207 & J121530.38+093505.9 & 23 & 39206 & 183.877 & 9.585 & 1 & 18.06 & 6 & 9.88 & 1 & 0.985 & 0.06 & 4 & 0.842 \\
NGC4206 & J121516.80+130126.3 & 24 & 39183 & 183.820 & 13.024 & 1 & 18.87 & 6 & 9.82 & 1 & 0.985 & -0.49 & 4 & 0.839 \\
NGC4144 & J120958.62+462726.6 & 25 & 38688 & 182.497 & 46.457 & 1 & 6.89 & 4 & 9.03 & 0 & 0.994 & -0.92 & 0 & 0.890 \\
NGC4713 & J124957.89+051840.7 & 27 & 43413 & 192.491 & 5.311 & 1 & 14.64 & 6 & 9.47 & 1 & 0.988 & -0.72 & 4 & 0.854 \\
NGC3109 & J100306.05-260933.2 & 28 & 29128 & 150.780 & -26.160 & 1 & 1.34 & 4 & 8.36 & 0 & 0.999 & -1.76 & 0 & 0.941 \\
\dots & \dots & \dots & \dots & \dots & \dots & \dots & \dots & \dots & \dots & \dots & \dots & \dots & \dots & \dots
\enddata
\tablecomments{510,072 entries. Available online (see text in Appendix~\ref{app:revised_sample}).}
\end{deluxetable}

\end{footnotesize}
\end{longrotatetable}

\end{changemargin}

\onecolumngrid
\section{Star formation rate distribution}\label{app:SFR_dist}

Following \cite{1979ApJ...232..352S}, the SFR functions of galaxies in the Local Volume are derived with an unbinned maximum-likelihood estimation. The SFR function of each morphological class of galaxies is modeled, as in \cite{2012ApJ...758..134S}, with a log-normal distribution centered on $\mu_{\rm pop}$ and of width $\sigma_{\rm pop}$, both expressed in dex. The number of objects between logarithmic SFRs $s$ and $s+\dd s$ is then

\begin{equation}
\frac{\dd N_{\rm true}}{\dd s} = N_{\rm pop} \times \mathcal{N}(s| \mu_{\rm pop} ,\sigma_{\rm pop}^2),
\end{equation}
where $N_{\rm pop}$ is the total number of objects in the population (both above and below sensitivity threshold). 

For a given distance, the minimum stellar mass above which a galaxy is observed is denoted as $M_{\rm 2MPZ}(d) = M_{\rm 2MPZ}(10\,{\rm Mpc}) \times (d/10\,{\rm Mpc})^2$, where $d$ is the luminosity distance. \cite{1979ApJ...232..352S} model this selection effect with a Fermi-Dirac step function. Without loss of generality, the selection effect is modeled in the present work with a Gaussian step function, $\mathcal{H}(m_*|m_{\rm 2MPZ}, \Delta m_*^2)$, where $\mathcal{H}$ is the Gaussian cumulative distribution function defined in Sec.~\ref{sec:sfr_function}, with width $\Delta m_*$ and mean $m_{\rm 2MPZ} = \log_{10} M_{\rm  2MPZ}(d) / M_{\rm ref}$, where $M_{\rm ref} = 10^9 M_\odot$.

Following the relations $s_\alpha = f(m_*)$ of slope $b$ with dispersion $\sigma_{m_*}$ established in Sec.~\ref{sec:sfr_calib}, the number of galaxies observed between $s_{\rm obs}$ and $s_{\rm obs} + \dd s_{\rm obs}$ at a given distance is, after integration over the true SFR, $s$,

\begin{equation}
\label{eq:numberL}
\frac{\dd N_{\rm obs}}{\dd s_{\rm obs}} = N_{\rm pop} \mathcal{H}(s_{\rm obs}|f(m_{\rm 2MPZ}), b^2\Delta m_*^2) \mathcal{N}(s_{\rm obs}|\mu_{\rm pop}, \sigma_{m_*}^2+\sigma_{\rm pop}^2),
\end{equation}

The likelihood to be maximized, $\mathcal{L} = \prod_{i=1}^{N}  \frac{1}{N_{\rm obs}} \frac{\dd N_{\rm obs}(s_i)}{\dd s_{\rm obs}}$, corresponds to the deviance

\begin{equation}
D = \sum_{i=1}^{N} \Bigg[\frac{(s_i - \mu_{\rm pop})^2}{\sigma_{m_*}^2 + \sigma_{\rm pop}^2} + 2 \ln\big(\sigma_{m_*}^2 + \sigma_{\rm pop}^2\big)- 2\ln\left(\frac{\mathcal{H}(s_{\rm i}|f(m_{\rm 2MPZ}), b^2\Delta m_*^2)}{\mathcal{H}(\mu_{\rm pop}|f(m_{\rm 2MPZ}), V_{\rm tot})}  \right) \Bigg],
\end{equation}
where $s_i$ is the logarithmic SFR of the galaxy indexed by $i$ and $V_{\rm tot} = \sigma_{\rm pop}^2 + \sigma_{m_*}^2 + b^2\Delta m_*^2$.

\begin{figure*}[ht!]
\epsscale{1.2}
\plotone{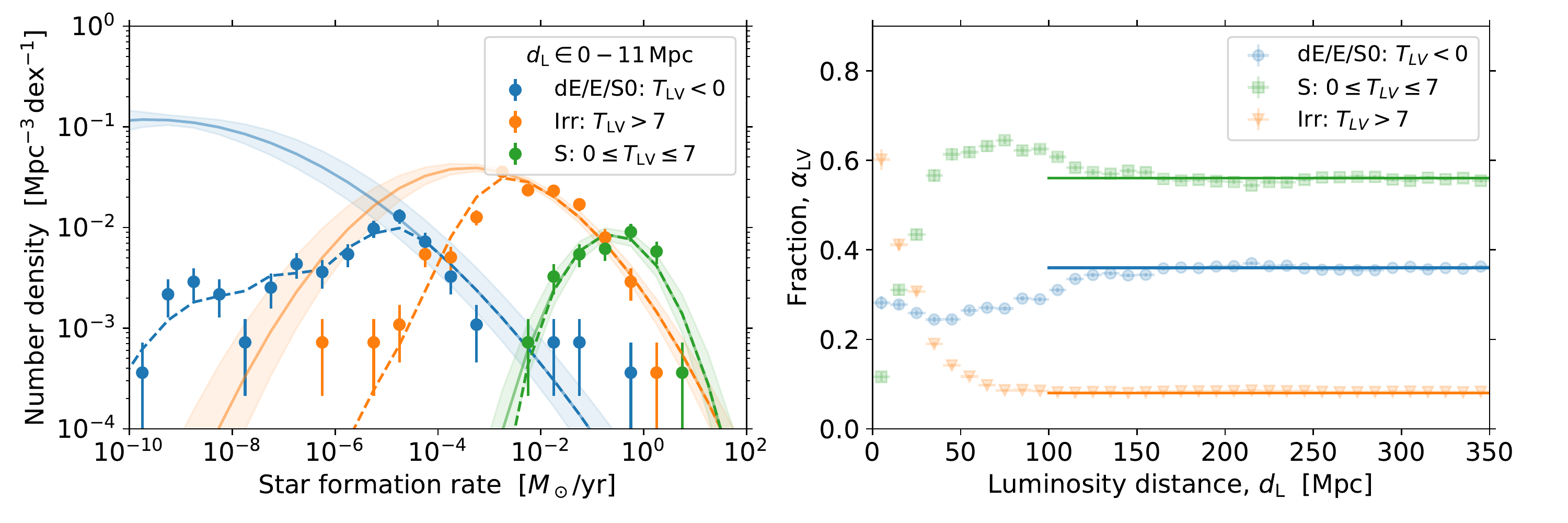}
\caption{\textit{Left}: The reconstructed SFR functions of the three morphological types of galaxies within 11\,Mpc. The observed counts are histogrammed in bins of 0.5\,dex. The dashed lines represent the best-fit model of the observed counts, while the solid lines show the true number counts expected with no sensitivity limit. \textit{Right}: The reconstructed fractions of the three morphological types of galaxies according to the LV classification in bins of 10\,Mpc. The fractions averaged between 100 and 350\,Mpc are indicated as solid lines.\label{fig:sfr_fun}\vspace{-0.3cm}}
\end{figure*}

The results obtained with the best-fit $\mu_{\rm pop}$, $\sigma_{\rm pop}$, $m_{\rm 2MPZ}$ and $\Delta m_*$ for the three morphological classes of galaxies with tabulated morphology above the 2MPZ sensitivity limit in the Local Volume sample are shown in Fig.~\ref{fig:sfr_fun}, left. 
The best fit parameters are $\mu({\rm S}) = -0.61 \pm 0.08 $ and $\sigma({\rm S}) = 0.63 \pm 0.07 $, $\mu({\rm Irr}) = -3.38 \pm 0.22 $ and $\sigma({\rm Irr}) = 1.33 \pm 0.11$, $\mu({\rm dE/E/S0}) = -9.6 \pm 1.0 $ and $\sigma({\rm dE/E/S0}) = 2.07 \pm 0.29 $ for spiral, irregular and elliptical galaxies, respectively. These results are obtained by fixing the sensitivity threshold at $\log_{10} M_{\rm 2MPZ}(10\,{\rm Mpc})/M_\odot = 7.54$, yielding a transition of width $\Delta m_* = 0.331 \pm 0.025$. Releasing the sensitivity threshold, now assumed to be a free parameter, yields a threshold value of $7.81 \pm 0.07$ and a width of $0.386 \pm 0.031$, as well as $\mu'({\rm S}) = -0.63 \pm 0.08$ and $\sigma'({\rm S}) = 0.64 \pm 0.07$, $\mu'({\rm Irr}) = -4.34 \pm 0.52$ and $\sigma'({\rm Irr}) = 1.53 \pm 0.16$, $\mu'({\rm dE/E/S0}) =  -10.6 \pm 1.3$ and $\sigma'({\rm dE/E/S0}) = 2.05 \pm 0.31$ for spiral, irregular and elliptical galaxies, respectively. The location of the SFR peak of spiral galaxies is robust against variations of the sensitivity threshold, while those of irregular and elliptical galaxies are smaller by about 1\,dex because their peaks lie in the region where the observed SFR function is affected by selection effects. For both a fixed and free sensitivity threshold, spirals, irregulars and ellipticals represent $55-60\%$, $45-40\%$ and less than $0.5\%$ of the total SFR in the Local Volume, respectively.

By fitting the UV/optical-based SFR function of galaxies down to a mass limit $\log_{10} M_*/M_\odot = 8$, \cite{2012ApJ...758..134S} identified two components peaking at $\mu_S = -0.70$ and $\mu_P = -2.82$ with widths  $\sigma_S = 0.72$ and $\sigma_P = 1.14$. The lack of a third component in the analysis of \cite{2012ApJ...758..134S} can be understood from a higher mass threshold with respect to the mass range accessible in the Local Volume. The widths of the two components identified by \cite{2012ApJ...758..134S} match within $1.5\,\sigma$ those of spiral and irregular galaxies estimated with a fixed threshold. A $0.1-0.6\,$dex difference is observed in the peak location, which could be understood from the use of different SFR tracers.

The most striking difference between the results presented in Fig.~\ref{fig:sfr_fun}, left, and those from \cite{2012ApJ...758..134S} lies in the relative abundance of irregular and spiral galaxies. The peak density of the latter is found to be lower than that of the former in the Local Volume (11\,Mpc) while the opposite holds for galaxies at $z<0.22$ (1,100\,Mpc) in the work of \cite{2012ApJ...758..134S}. A possible solution to this apparent discrepancy lies in the sub- and super-linear dependence of SFR on stellar mass of spiral and irregular galaxies, respectively. For a mass-limited sample, the population of galaxies at higher redshifts is expected to be dominated by spirals rather than irregulars. This is in line with the observed proportions of galaxies with a morphological type tabulated in HyperLEDA, as shown Fig.~\ref{fig:sfr_fun}, right. Galaxies are distributed, following Eq.~\ref{eq:part_morph}, as ${\sim}\,10\%$ spirals, ${\sim}\,60\%$ irregulars and ${\sim}\,30\%$ ellipticals out to 10\,Mpc, and as  ${\sim}\,55\%$ spirals, ${\sim}\,10\%$ irregulars and ${\sim}\,35\%$ ellipticals between 100 and 350\,Mpc. The SFR incompleteness correction factors in Eq.~\ref{eq:cS} are based on the average fractions illustrated as solid lines in Fig.~\ref{fig:sfr_fun}, right. The assumption of a constant partitioning across the three morphological types, although breaking down at low distances, results in an uncertainty on SFR incompleteness in the Local Volume estimated to be lower than 5\%.

\section{3D visualizations}\label{app:3d}

\begin{figure}[h]
    \epsscale{1.2}
    \plotone{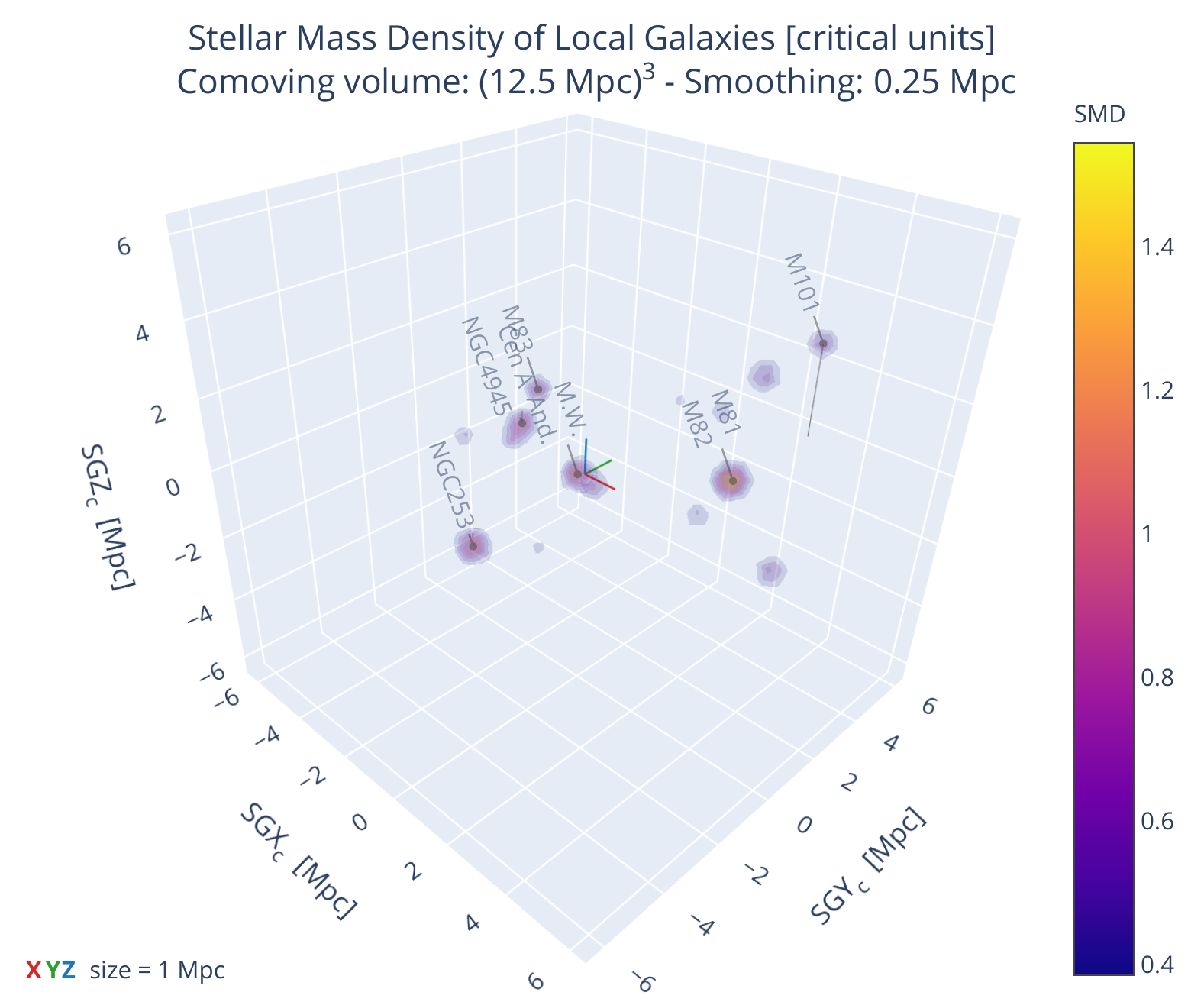}
    \caption{The SMD in a 12.5\,Mpc-side cube, in comoving supergalactic coordinates, smoothed on a comoving scale of 0.25\,Mpc. See Fig.~\ref{fig:3D}. Interactive version available online.
    \label{fig:3D_a}}
    \end{figure}

\begin{figure}[h]
    \epsscale{1.2}
    \plotone{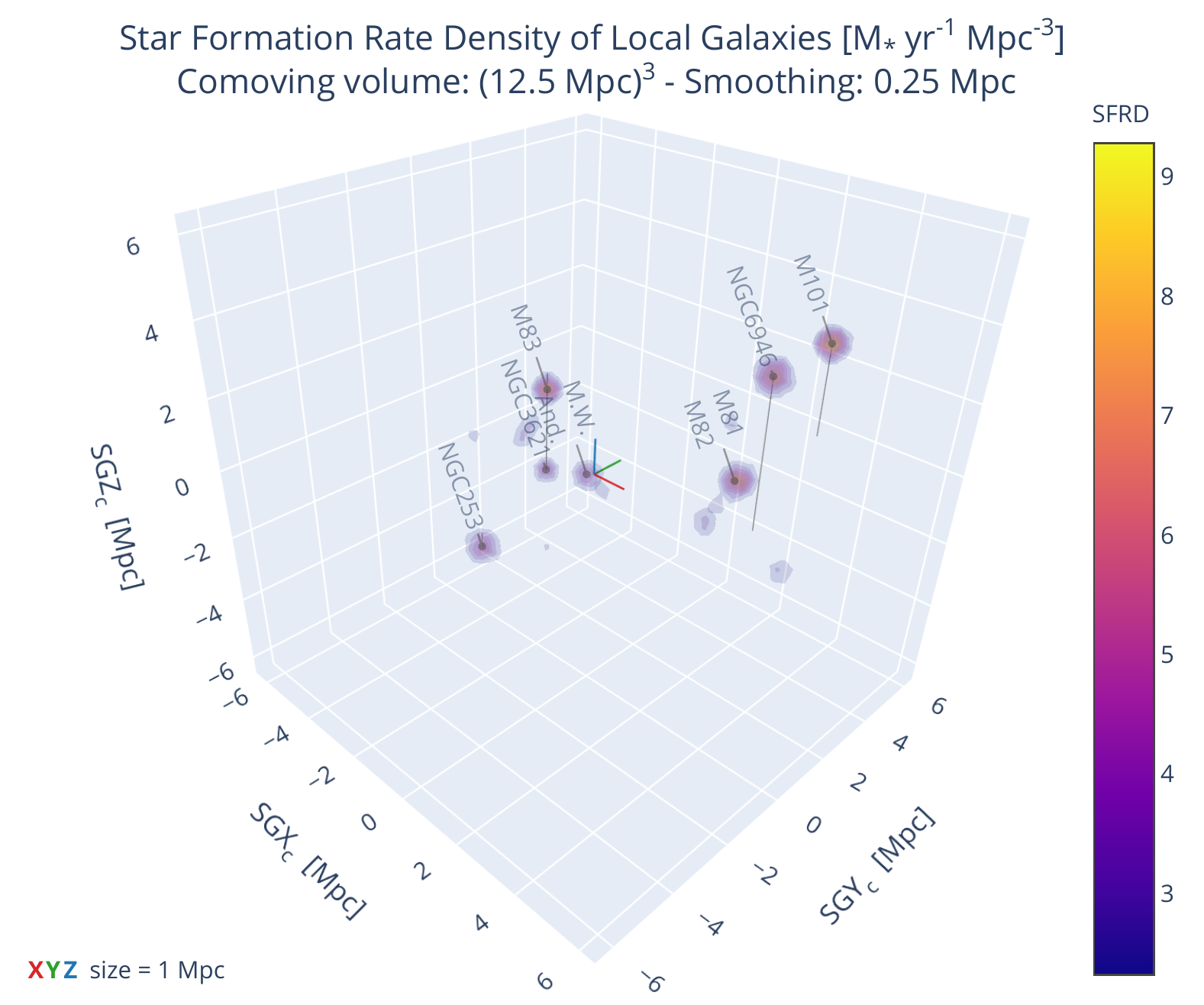}
    \caption{The SFRD in a 12.5\,Mpc-side cube, in comoving supergalactic coordinates, smoothed on a comoving scale of 0.25\,Mpc. See Fig.~\ref{fig:3D}. Interactive version available online.
    \label{fig:3D_b}}
    \end{figure}

\begin{figure}[h]
    \epsscale{1.2}
    \plotone{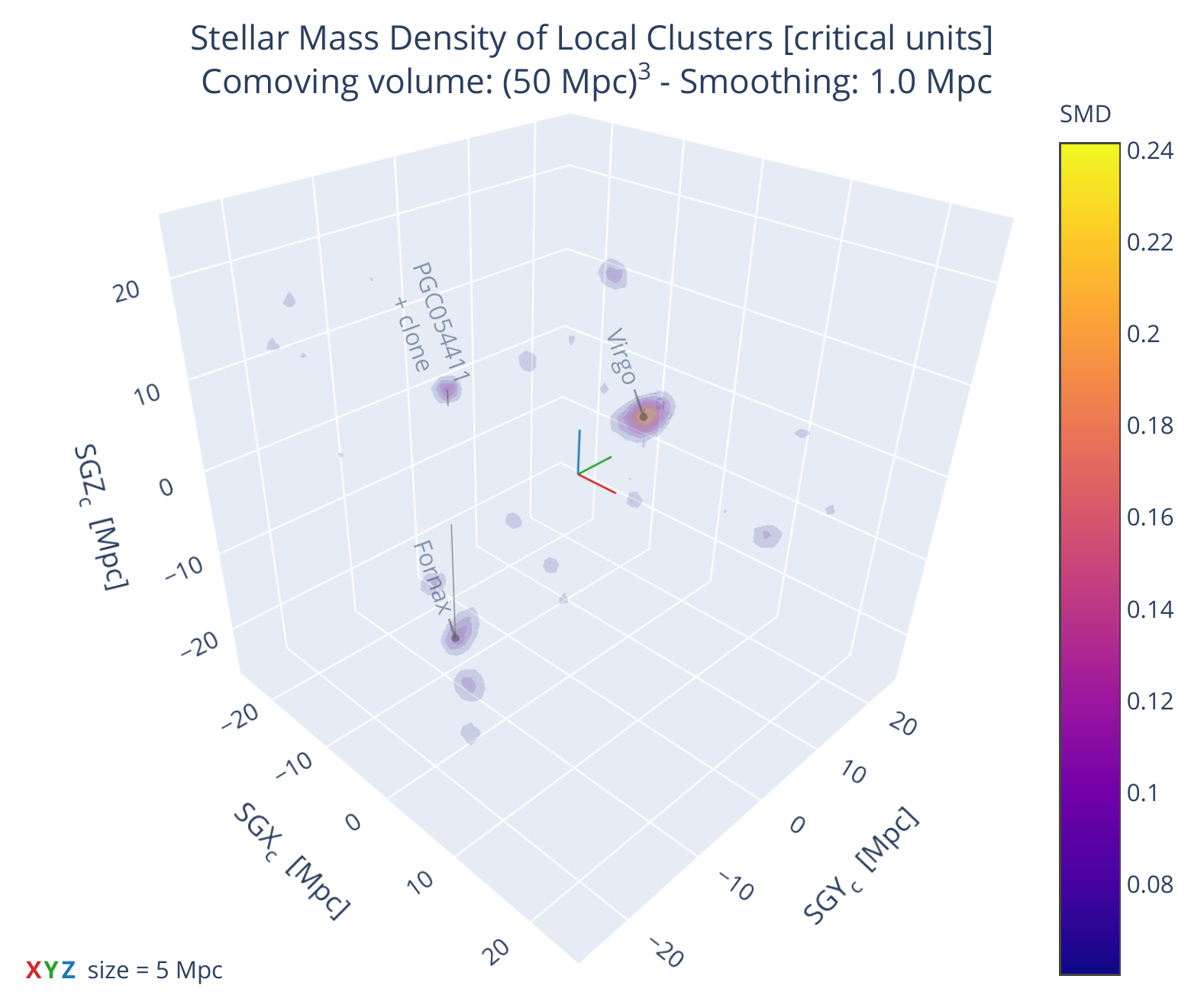}
    \caption{The SMD in a 50\,Mpc-side cube, in comoving supergalactic coordinates, smoothed on a comoving scale of 1\,Mpc. See Fig.~\ref{fig:3D}. Interactive version available online.
    \label{fig:3D_c}}
    \end{figure}

\begin{figure}[h]
    \epsscale{1.2}
    \plotone{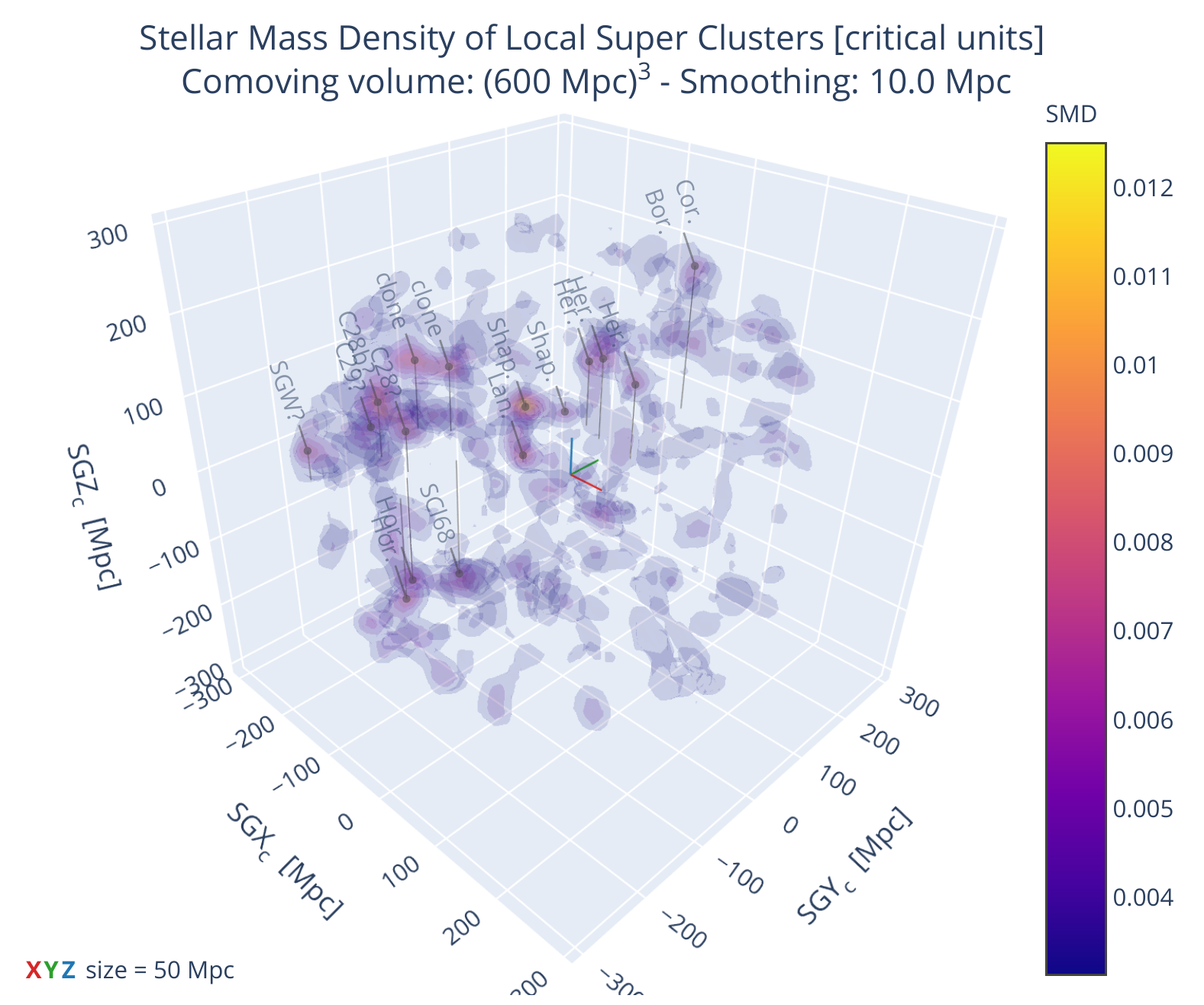}
    \caption{The SMD in a 600\,Mpc-side cube, in comoving supergalactic coordinates, smoothed on a comoving scale of 10\,Mpc. See Fig.~\ref{fig:3D}. Interactive version available online.
    \label{fig:3D_e}}
    \end{figure}

\begin{figure}[h]
    \epsscale{1.2}
    \plotone{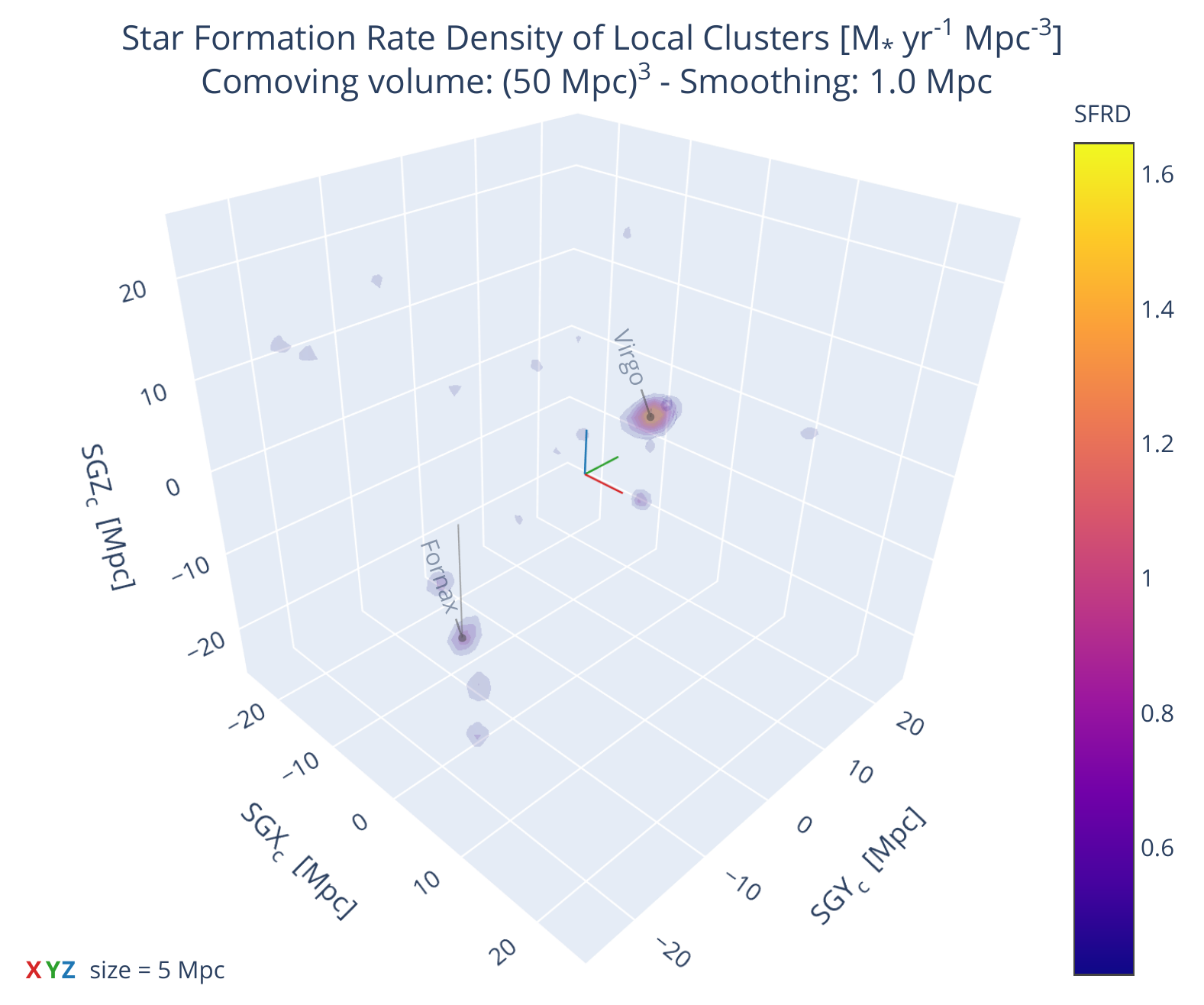}
    \caption{The SFRD in a 50\,Mpc-side cube, in comoving supergalactic coordinates, smoothed on a comoving scale of 1\,Mpc. See Fig.~\ref{fig:3D}. Interactive version available online.
    \label{fig:3D_d}}
    \end{figure}

\begin{figure}[h]
    \epsscale{1.2}
    \plotone{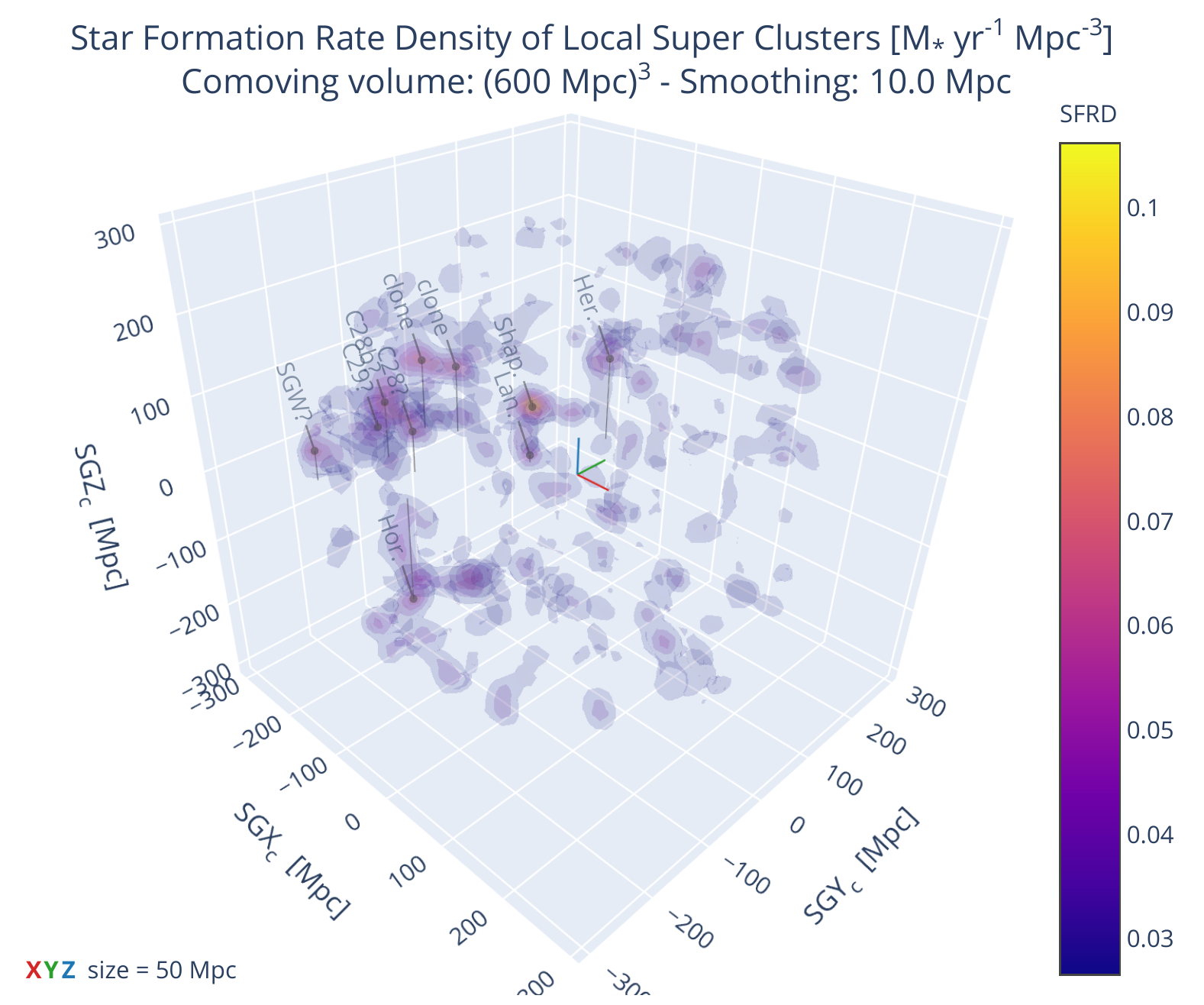}
    \caption{The SFRD in a 600\,Mpc-side cube, in comoving supergalactic coordinates, smoothed on a comoving scale of 10\,Mpc. See Fig.~\ref{fig:3D}. Interactive version available online.
    \label{fig:3D_f}}
    \end{figure}

The SMD and SFRD are displayed in Figs.~\ref{fig:3D_a}-\ref{fig:3D_f} on scales ranging from that of the Local Sheet out to distant superclusters. Acronyms denote the Milky Way (\texttt{M.W.}) and Andromeda (\texttt{And.}) in Figs.~\ref{fig:3D_a} and \ref{fig:3D_b}, and they denote in Figs.~\ref{fig:3D_e} and \ref{fig:3D_f} the Laniakea (\texttt{Lan.}), Shapley (\texttt{Shap.}), Hercules (\texttt{Her.}), Corona Borealis  (\texttt{Cor.\ Bor.}), and Horologium (\texttt{Hor.})\ superclusters, as well as supercluster 68 (\texttt{SCl68}), the Southern Great Wall (\texttt{SGW?}) and the C29/C28 (\texttt{C29?}, \texttt{C28?}, \texttt{C28b?}) overdensities identified in \cite{2006MNRAS.373...45E}. 

\twocolumngrid

\clearpage
\bibliography{sm_sfr_1Gyr}{}
\bibliographystyle{aasjournal}

\end{document}